\documentclass[aps,prb,twocolumn,groupedaddress,showpacs,showkeys]{revtex4-1}
\usepackage{amsmath}
\usepackage{float}
\usepackage{amssymb}
\usepackage{graphicx}
\usepackage{epstopdf}
\usepackage{esint}
\usepackage{dutchcal}

\usepackage[usenames,dvipsnames]{color}

\begin{document}

\title{Nanoscale Plasmonic Slot Waveguides for Enhanced Raman Spectroscopy}

\author{Herman M. K. Wong$^1$, Mohsen Kamandar Dezfouli$^2$, Lu Sun$^1$,  Stephen Hughes$^2$, and Amr S. Helmy$^1$}
\email{a.helmy@utoronto.ca}

\affiliation{$^1$Department of Electrical and Computer Engineering, University of Toronto, Toronto, ON M5S 3G4, Canada}
\affiliation{$^2$Department of Physics, Engineering Physics and Astronomy, Queen's University, Kingston, ON K7L 3N6, Canada}

\date{\today}

\begin{abstract}
                We theoretically investigate several types of plasmonic slot waveguides for enhancing the measured signal in Raman spectroscopy, which is a consequence of electric field and Purcell factor enhancements, as well as an increase in light-matter interaction volume and the Raman signal collection efficiency. An intuitive methodology is presented for calculating the accumulated Raman enhancement factor of an ensemble of molecules in waveguide sensing, which exploits an analytical photon Green function expansion in terms of the waveguide normal modes, and we combine this with a quantum optics formalism of the molecule-waveguide interaction to model Raman scattering. We subsequently show how integrated plasmonic slot waveguides can attain significantly higher Raman enhancement factors: $\sim$5.3$\times$ compared to optofluidic fibers and $\sim$3.7$\times$ compared to planar integrated dielectric waveguides, with a device size and thus analyte volume of at least three-orders of magnitude less.
%Although conventional surface enhanced Raman spectroscopy (SERS) can attain %comparable EFs, plasmonic slot waveguides have other advantages including %the versatility for on-chip integration with other components, whereas SERS %requires bulky equipment such as a confocal microscope. Furthermore, 
We also provide a comprehensive comparison between the different types of plasmonic slot waveguides based on the important figures-of-merit, and determine the optimal approaches to maximize Raman enhancement.
\end{abstract}

 \maketitle

% ===============================================================================

\section{Introduction}
\label{sec:Intro}

Due to the confluence of technologies into portable devices such as smartphones, wearable computing, and generally the connected nodes of the ``Internet of Things'' \cite{Swan2012,AlFuqaha2015}, there is an emerging need for ultra-compact sensors with high specificity and label-free detection of a wide range of chemical and biological molecules, for which Raman spectroscopy serves as an ideal technology \cite{Baraldi2008,Patel2010,Efremov2008,Downes2010,Tanaka2006,Chalmers2012,Sarmiento2008}. However, spontaneous Raman scattering, that can be employed in a portable setting, is exceedingly weak \cite{Kneipp2006}, and a major challenge is the effective separation of the Raman scattered photons from the intense Rayleigh scattering of the pump light.

A formidable approach to boost the signal of Raman scattered light is surface enhanced Raman spectroscopy (SERS), which often makes use of metal nanostructures to enhance local electromagnetic (EM) fields and the photonic local density of states (LDOS) by the creation of plasmonic hot-spots. Thus, metal-assisted SERS effectively increases the Raman scattering cross-section of each molecule. Experimentally, impressive SERS enhancement factors of up to $10^{14}$ have been observed in composites of metal nanoparticles \cite{Kneipp2006}, which includes dimers and small aggregates formed by metal nanoparticles \cite{Xu2000}, as well as fractal types of nanostructures \cite{Li2003}. The main limitation of SERS is that bare substrates typically require surface functionalization to bring the analyte of interest close enough to the nanostructured metal surface (a few nanometers) for effective sensing \cite{Stiles2008,Carron1992}, and there are several challenges that limit the use of bare SERS substrates \cite{Stiles2008}, including the instability due to oxidation of the metal surface \cite{VonRaben1984,Fornasiero1987}.

A complementary strategy to enhance Raman spectroscopy is by increasing the effective light-matter interaction volume for Raman scattering, which is typically achieved by inducing the interaction of light with analyte molecules along the length of a coupled waveguide (WG). Simultaneously, the WG can also offer higher collection efficiency of Raman scattered light compared to collection in free space. In this regard, hollow core WGs that confine both light and analyte fluid within the same volume, which in turn leads to high coupling efficiencies for both pump and Raman scattered light via the large numerical aperture (NA), such as teflon capillary tubes (TCTs) \cite{Altkorn1999} and hollow core photonic crystal fibers (HC-PCFs) \cite{Mak2013,Eftekhari2011}, have been demonstrated to provide very high Raman signal enhancement factors (up to $10^6$ compared to using a cuvette). Integrated WGs ``on-chip'' have also been investigated for increased light-matter interaction volume to enhance Raman spectroscopy \cite{Dhakal2014,Dhakal2015,Dhakal2016,Evans2016,Holmstrom2016}, with much reduced device sizes and the simplicity of placing a droplet of analyte liquid directly on top of the chip.

While both SERS and WG-based Raman spectroscopy serve to significantly enhance the retrieved Raman signal, the fundamental difference between these two techniques is that SERS enhances the intrinsic Raman scattered light intensity from each molecule, whereas the WG configuration increases the number of molecules that interact with the pump light and thus undergo Raman scattering. Although the maximal enhancement achieved by SERS is extremely high, the distribution and generation of hotspots on a SERS substrate is generally quite random and unstable \cite{Mak2013}, and thus the Raman enhancement is not very   repeatable. Moreover, the substrate surface must typically be chemically treated to facilitate effective binding of analyte molecules \cite{Stiles2008,Carron1992}. The volume occupied by each hotspot is also very small, and the number of hotspots is limited, so the overall volume of analyte that can be interrogated is not much higher than by direct illumination of the sample contained in a cuvette. On the other hand, while having the fluid sample sit on top of a planar WG or fill a hollow core fiber increases the light-matter interaction volume, the Raman scattered light intensity from each molecule is not enhanced by much beyond when it is excited in free space. 

In this work, we theoretically investigate nanoscale plasmonic slot WG designs for their application to Raman spectroscopy in a regime where signal enhancement can be engineered through two key mechanisms: first, by exploiting the well-known local electric-field ($\textbf{E}$-field) and LDOS enhancement near metal surfaces as in traditional SERS; and second, by increasing the light-matter interaction volume and Raman signal collection efficiency via light confinement and WG propagation. Using a plasmonic slot of only tens of nanometers wide, in which the analyte fluid is filled, the molecules under investigation are already in close proximity to the metal surfaces (i.e., sidewalls of the slot), and thus the requirement for chemical surface treatment can be eliminated. Moreover, the slot WG structure can be precisely controlled during fabrication, which means the guided plasmonic mode can be excited repeatably. To model these complex geometries and interactions, we combine a first-principles electromagnetics and quantum optics approach, starting with the $\textbf{E}$-field Green function of the WG in terms of normal modes \cite{MangaRao,Cano2013}, and then generalizing a recently developed quantum optomechanical theory to obtain the Raman enhancement \cite{Dezfouli2017,Wong2017}. This serves to rigorously determine the advantages and application niche of our proposed plasmonic slot WGs in comparison to previous techniques for Raman enhancement, and allows one to quickly assess the important figures-of-merit in a semi-analytical and intuitive way.

The layout of the rest of our paper is as follows. In Section\,\ref{sec:Formalism}, we define the important figures-of-merit to characterize and compare WGs for their performance in Raman enhancement, and introduce the main theoretical formalism that allows one to calculate these figures-of-merit. In Section\,\ref{sec:Results}, we consider several types of plasmonic slot WGs (and also a dielectric slot WG), and present our results for different figures-of-merit of these structures. The key characteristics of Raman enhancement using WGs that include different excitation and Raman signal collection configurations are highlighted. Finally, we conclude in Section\,\ref{sec:Conclusions}, with a discussion on the most promising strategy to engineer plasmonic slot WGs for increased Raman enhancement, the practical limitations of implementing slot WGs for Raman spectroscopy, and other advantages of employing plasmonic slot WGs in comparison to more conventional techniques such as hollow core fibers and SERS on a planar substrate. In addition, we include seven appendices. In Appendix A, we give the calculation of the energy velocity and waveguide mode normalization scheme. Appendix B shows a comparison of our waveguide Green function formulation with the one given in Ref.\,\citenum{Cano2013}. In Appendix C, the derivation of the $\textbf{E}$-field enhancement factor is shown. This is followed by the specification of the integration geometries used to calculate the Raman scattered power using WGs as well as in free space, shown in Appendix D. In Appendix E, we show the derivation of the Raman enhancement factor that captures the effects of both SERS-type and light-matter interaction volume enhancements. Appendix F gives the results of Raman enhancement for different modes of the Rhodamine 6G molecule, which we utilize in this work as a representative example. Lastly, Appendix G contains additional figures showing results referred to by the main text.

% =============================================================================================

\section{Raman Enhancement in Waveguides: Theoretical Formalism}
\label{sec:Formalism}

\begin{figure*}
        \centering
        \includegraphics[width=\textwidth]{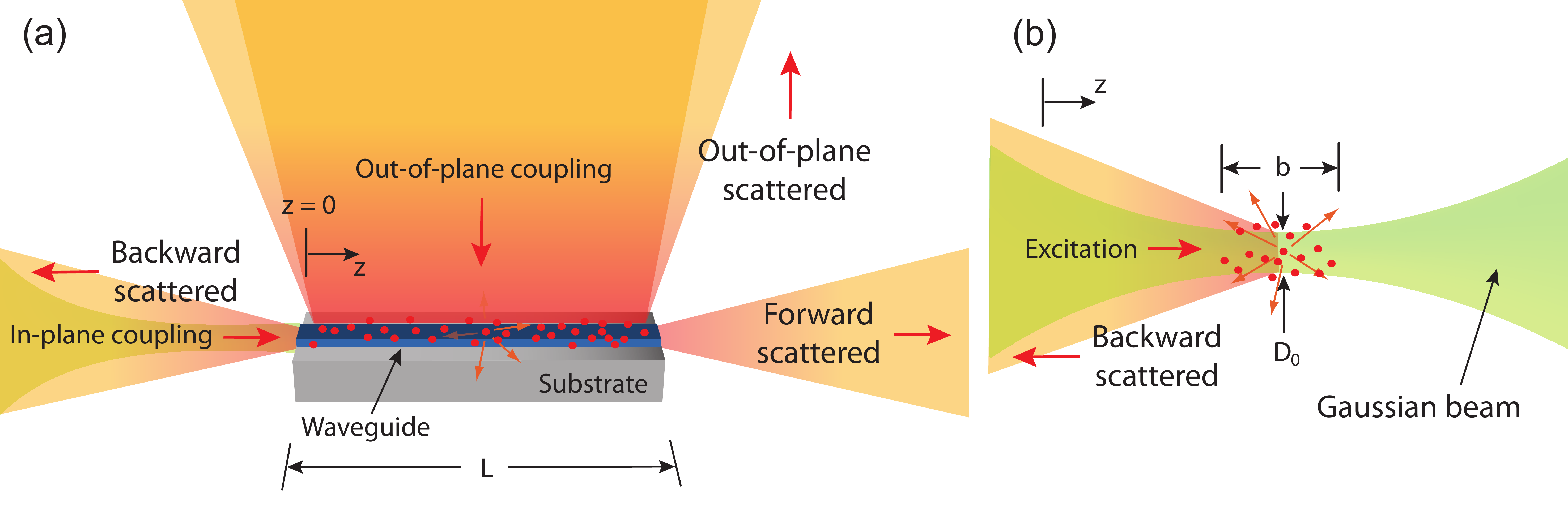}
        \caption{(a) Schematic of different configurations of laser excitation and Raman scattered light collection with respect to the WG of length $L$ oriented in the light guiding $z$-direction. Excitation is either (A) in-plane via coupling to the WG mode or (B) out-of-plane from the top. Collection is either in the (I) forward, (II) backward or (III) out-of-plane scattered direction. (b) Schematic of laser excitation and Raman scattered light collection in free space. Excitation is in the forward direction and Raman scattered light is collected in the backward direction. The Gaussian beam is propagating in the $z$-direction with waist diameter $D_0$ and depth of focus $b$.}
        \label{fig:exp_config}
\end{figure*}

%Raman enhancement using waveguide structures  proposed in this work is fundamentally %attributed to two different enhancement mechanisms: (1) each molecule experiences %an enhancement due to the increased excitation $\textbf{E}$-field intensity %as well as the enhanced LDOS, and (2) the enhancement due to the increased %light-matter interaction volume via light propagation along the waveguide. 
We first define the main figure-of-merit of Raman enhancement for a single molecule (SM), which we denote as the single-molecule enhancement factor (SMEF). For Raman spectroscopy of a bulk-like sample (with an ensemble of molecules), an alternative spatially averaged Raman enhancement factor (AEF) is defined to also capture the effects of increased light-matter interaction volume using an ensemble of molecules, which allows us to compute the volume enhancement factor (VEF) for a specific device length, and compare it to a reference Gaussian beam. Below, we adapt a recently developed quantum optics approach to model the Raman spectrum of a molecule within a general medium, which was previously applied to describe SERS using plasmonic nanoresonators \cite{Dezfouli2017,Wong2017}. Here, we extend this approach to WGs, by utilizing the photon Green function, conveniently computed using a
semi-analytical normal mode theory  \cite{MangaRao,Cano2013}.

\subsection{Raman Enhancement of a Single Molecule}
\label{subsec:SMEF}

A Raman active molecule in the vicinity of a WG can be excited either by (A) in-plane coupling to the 2-D WG mode or (B) out-of-plane excitation [see Fig.\,\ref{fig:exp_config}(a)]. Upon excitation, the Raman scattered light can either be coupled to the WG mode in the (I) forward (collected at the end facet of the WG, $z = L$), (II) backward (collected at the input facet of the WG, $z = 0$), or (III) out-of-plane direction via radiation modes above the light line [Fig.\,\ref{fig:exp_config}(a)]. The detected Raman scattered power [W] from a single molecule, in the vicinity of the WG that we denote $P^\textnormal{SM}$, thus depends on the position of the molecule $\textbf{r}_m$ (determined by the location within the 2-D cross-section $\boldsymbol{\rho}_m$, and the position along the length of the WG $z_m$ away from the input facet, such that $\textbf{r}_m = [\boldsymbol{\rho}_m,z_m]$), the WG length $L$, and also the specific configuration of excitation of the molecule and collection of the Raman scattered light [Fig.\,\ref{fig:exp_config}(a)]. Due to the enhancement of the excitation field and also the LDOS (yielding a Purcell effect or spontaneous emission enhancement) that a molecule can experience in the vicinity of the WG, $P^\textnormal{SM}$ can be enhanced compared to the Raman scattered power from a molecule in free space (or homogeneous background) $P^{\rm SM}_0$. In this way, the SMEF is defined as
\begin{equation}
\label{eq:SMEF}
{\rm SMEF} = \frac{P^{\textnormal{SM}}(\mathbf{r}_m)}{P_0^{\rm SM}},
\end{equation}
which is different for each Raman mode as well as whether it is for the Stokes or anti-Stokes resonance. 

The Raman scattered power from a single molecule, integrated over a detector area defined from ${\bf r}_\textnormal{D}$, can be calculated from
\begin{equation}
\label{eq:SERS_power}
P^{\textnormal{SM}}(\mathbf{r}_m) = \frac{\varepsilon_0 c n_\textnormal{B}}{2} \int_{\Delta\omega}\int_{\mathcal{A}_\textnormal{D}}S(\mathbf{r}_\textnormal{D},\mathbf{r}_m,\omega)d\mathbf{r}_\textnormal{D}\,d\omega,
\end{equation}
where $n_\textnormal{B} = \sqrt{\varepsilon_\textnormal{B}}$ is the refractive index of the background medium ($\varepsilon_\textnormal{B}$ is the relative permittivity or dielectric constant), and the integration is over both the detection area $\mathcal{A}_\textnormal{D}$ and the bandwidth of the Raman line of interest $\Delta\omega$. The detection area $\mathcal{A}_\textnormal{D}$ and the integration geometry for Eq.\,\eqref{eq:SERS_power} are dependent on whether the Raman scattered power of a molecule in the vicinity of a WG ($P^\textnormal{SM}$) or in free space ($P^\textnormal{SM}_0$) is being calculated; further details are given in Appendix \ref{Appendix:Calc_G}. Using a quantum optomechanical theory of SERS and its detection \cite{Dezfouli2017,Wong2017}, the Raman spectrum is expressed in terms of the $\textbf{E}$-field operator $\hat{\textbf{E}}$ by $S(\textbf{r}_\textnormal{D},\textbf{r}_m,\omega) = \langle\hat{\textbf{E}}^\dagger(\textbf{r}_\textnormal{D},\textbf{r}_\textnormal{m};\omega)\cdot\hat{\textbf{E}}(\textbf{r}_\textnormal{D},\textbf{r}_\textnormal{m};\omega)\rangle$ and can be calculated as
\begin{equation}
\label{eq:Scat_spect}
S(\mathbf{r}_\textnormal{D},\mathbf{r}_m,\omega) = A(\mathbf{r}_m,\omega_\textnormal{L},\mathbf{E}_0)S_0(\mathbf{r}_m,\omega)|\mathbf{G}(\mathbf{r}_\textnormal{D},\mathbf{r}_m;\omega) \cdot \mathbf{n}|^2,
\end{equation}
where $\textbf{n}$ specifies the direction of dipole orientation (direction of dominant Raman tensor element of molecule). The incident field enhancement is described through $A(\mathbf{r}_m,\omega_\textnormal{L},\mathbf{E}_0)$, the emitted spectrum from the molecule is described by $S_0(\mathbf{r}_m,\omega)$, and the propagation to the detection region is contained within the two space point Green function, $\mathbf{G}(\mathbf{r}_\textnormal{D},\mathbf{r}_m;\omega)$, in which $\textbf{r}_m$ and $\textbf{r}_\textnormal{D}$ are the molecule and detection locations, respectively. This expression is quite general, and different photonic media can have different Green functions, e.g., 
in free space or a homogeneous medium (background index $n_\textnormal{B}$), the Green function is known analytically
% and given by ${\bf G}_{0, ii} =\omega^3 n_\textnormal{B}/(6\pi c^3$
\cite{Novotny2012}.

For a non-periodic WG that is translationally invariant along the WG direction, having effectively a 1-D-like propagation in the $z$-direction, the single mode Green function  is \cite{MangaRao,Cano2013}
\begin{align}
& \mathbf{G}_\textnormal{WG}(\textbf{r},\mathbf{r}^\prime;\omega) = \frac{i\omega}{2\upsilon_\textnormal{E}}\Big[\Theta(z - z^\prime)\mathbf{e}_{k_\omega}(\boldsymbol{\rho})\mathbf{e}_{k_\omega}^*(\boldsymbol{\rho}^\prime)e^{i\tilde{k}(z-z^\prime)} \nonumber \\  
& + \Theta(z^\prime - z)\mathbf{e}_{k_\omega}^*(\boldsymbol{\rho})\mathbf{e}_{k_\omega}(\boldsymbol{\rho}^\prime)e^{i\tilde{k}(z^\prime-z)}\Big],
\label{eq:Green_function} 
\end{align}
where the first term  represents forward propagation, and the second term represents backward propagation; in addition, $\Theta$ is the unit step function, $\tilde{k} = k + i\kappa$ is the complex propagation constant where $\kappa$ incorporates the propagation losses, $\textbf{e}_{k_\omega}(\boldsymbol{\rho})$ is the normalized mode with the propagation constant $k(\omega)$ that we obtain using the commercial mode solver from Lumerical \cite{Lumerical}, and $\upsilon_\textnormal{E}$ is the frequency-dependent energy velocity that we employ instead of the group velocity, $v_g$, commonly used in lossless media. The energy velocity $\upsilon_\textnormal{E}$ is dependent on both the time-averaged power flow $\langle S_z \rangle$ as well as the time-averaged energy per length $\langle {\cal W} \rangle$, which in turn is dependent on the energy density distribution ${\cal w}(\boldsymbol{\rho})$ of the WG mode. Details about the calculation of the energy velocity $\upsilon_\textnormal{E}$ (including $\langle S_z \rangle$, $\langle {\cal W} \rangle$, and ${\cal w}$) and the normalization scheme to obtain $\textbf{e}_{k_\omega}(\boldsymbol{\rho})$ are given in Appendix \ref{Appendix:energy_velocity}. The total system Green function can then be written as the sum of the WG mode contribution (which dominates) and a background term (e.g., from coupling to modes above the light line), ${\bf G} \approx {\bf G}_{\rm WG}+{\bf G}_{\rm B}$, and will be used to calculate the Raman scattered spectrum of the molecules. %For the waveguides in this study, it is reasonable to assume ${\bf G}_{\rm B}=
%{\bf G}_{0} $, and the WG contribution will be the dominant term of interest.

For an incident field ${\bf E}_0$, the field enhancement of the induced Raman dipole is given by \cite{Dezfouli2017}
\begin{equation}
\label{eq:Prefactor}
A(\textbf{r}_m,\omega_\textnormal{L},\mathbf{E}_\textnormal{0}) = \frac{\hbar R_{nn}^2 |\eta(z_m)|^2 |\mathbf{n} \cdot \mathbf{E}_\textnormal{0}|^2}{2\omega_m\varepsilon_0^2},
\end{equation} 
where $R_{nn}$ is the dominant Raman tensor element of the molecule, $\omega_m$ is the Raman active oscillation frequency, $\textbf{n}$ points in the direction of the dominant Raman tensor element, and $|\eta|$ ($\textbf{r}_m$ dependence is implied) is the plasmonic field enhancement factor. Significant enhancement can be attained when excitation is by in-plane coupling to the WG mode (A), whereas there is essentially no $\textbf{E}$-field enhancement (i.e., $|\eta| = 1$) when excitation is from out-of-plane (B), as it does not couple to WG modes below the light line. For plasmonic nanoresonators, the enhancement factor can be calculated using a Dyson equation technique; but for WGs, a different approach is needed. As shown in Appendix\,\ref{Appendix:field_enhancement}, the field enhancement factor for molecules at different locations along the WG, $z_m$, can be written as:
\begin{equation}
\label{eq:field_enhancement}
|\eta(z_m)|^2 = \frac{c}{n_B\upsilon_\textnormal{E}}\frac{\mathcal{A}_0}{\mathcal{A}_\textnormal{WG}}e^{-2\kappa z_m},
\end{equation}
where $\mathcal{A}_0$ is the effective Gaussian beam area of the input laser in free space, $\mathcal{A}_{\rm WG}$ is the effective WG mode area, and the exponential factor accounts for propagation losses.

The emission spectral function can be written as $S_0(\textbf{r}_m,\omega) = S_0^{\textnormal{st}}({\bf r}_m,\omega) + S_0^{\textnormal{as}}({\bf r}_m,\omega)$, in which the Stokes emission is \cite{Dezfouli2017}
\begin{align}
& S_0^{\textnormal{st}}({\bf r}_m,\omega) = \label{eq:S0_st} \\
& {\rm Re}\left\{\frac{i[\gamma_m(\bar{n}^{\textnormal{th}} + 1) + J_{\textnormal{ph}}({\bf r}_m,\omega_L + \omega_m)]}{[\omega - (\omega_L - \omega_m) + i(\gamma_m +\Delta J_{\textnormal{ph}})](\gamma_m + \Delta J_{\textnormal{ph}})}\right\}, \nonumber
\end{align} 
and the anti-Stokes emission is
\begin{align}
& S_0^{\textnormal{as}}({\bf r}_m,\omega) = \label{eq:S0_as} \\
& {\rm Re}\left\{\frac{i[\gamma_m\bar{n}^{\textnormal{th}} + J_{\textnormal{ph}}({\bf r}_m,\omega_L - \omega_m)]}{[\omega - (\omega_L + \omega_m) + i(\gamma_m +\Delta J_{\textnormal{ph}})](\gamma_m + \Delta J_{\textnormal{ph}})}\right\} \nonumber,
%\label{eq:S0}
\end{align} 
where $\gamma_m$ is the decay rate of the Raman vibrational mode, and $\bar{n}^{\textnormal{th}}=\left(e^{\hbar\omega_m/k_B T} - 1\right)^{-1}$ is the thermal population of the Raman vibrational mode at room temperature, $T=293\,{\rm K}$. The photonic spectral function, $J_{\textnormal{ph}}$, describes the plasmonic-induced Raman scattering rate beyond that achievable through thermal population, which depends on the projected LDOS,
\begin{align}
J_{\textnormal{ph}}({\bf r}_m,\omega) =
\frac{R_{nn}^2|\eta(z_m)|^2 |\mathbf{n} \cdot \mathbf{E}_0|^2}{2\varepsilon_0\omega_m}\textnormal{Im}\{\mathbf{G}_{nn}(\mathbf{r}_m,\mathbf{r}_m;\omega)\},
\label{eq:Jph}
\end{align}
and for convenience, in Eqs.\,\eqref{eq:S0_st} and (\ref{eq:S0_as}), we have defined $\Delta J_{\textnormal{ph}} = J_{\textnormal{ph}}({\bf r}_m,\omega_L + \omega_m) - J_{\textnormal{ph}}({\bf r}_m,\omega_L - \omega_m)$. Note that $J_\textnormal{ph}$ will affect the system spectrum when it is comparable to $\gamma_m\bar{n}^{\textnormal{th}}$, which is normally achieved in the nonlinear regime of high pump excitations.

The last term of Eq.\,\eqref{eq:Scat_spect}, $|\mathbf{G}(\mathbf{r}_\textnormal{D},\mathbf{r}_m;\omega)\cdot\mathbf{n}|^2$, accounts for the enhancement of the Raman scattered field and the appropriate propagation effects from the molecule to the detector via the system Green function $\mathbf{G}(\mathbf{r}_\textnormal{D},\mathbf{r}_m;\omega)$. For the cases of Raman scattered light coupled to either the forward (I) or backward (II) propagating modes, the WG modal contribution $\textbf{G}_\textnormal{WG}$ as given in Eq.\,\eqref{eq:Green_function} is utilized. However, the free space Green function $\textbf{G}_{\rm B}$ is employed when out-of-plane (III) Raman scattered light is collected, which we have numerically verified to be a good approximation for the WGs under consideration. Note that $\textbf{G}_\textnormal{B}$ is also used for the calculation of Raman scattered power in free space, $P_0^\textnormal{SM}$.

\subsection{Purcell Factor and Waveguide Beta Factor}
\label{subsec:PF}

Engineered nanostructures can be utilized for nanoscale confinement of light as well as highly enhancing light-matter interaction. Examples of these structures include photonic crystals \cite{Lodahl2004,Yao2010,Hughes2004}, metamaterials \cite{Jacob2010,Noginov2010}, slow light WGs \cite{Yao2009}, plasmonic nanostructures \cite{Anger2006,Frimmer2013}, and plasmonic WGs \cite{Kress2015,Burmudez2015,Perera2013,Barthes2013,Chen2010,Chen2010_2,Jun2008}, which allow for enhancement of the LDOS of embedded quantum emitters, and thus increase in their spontaneous emission (SE) rates via the Purcell effect. Such enhancement benefits a number of applications including nonlinear optics \cite{Kauranen2012}, chemical sensing \cite{Homola1999,Kneipp1997}, high-resolution imaging \cite{Zhang2013}, energy harvesting \cite{Atwater2010,Yu2010}, and single photon sources \cite{Chang2007}. For many of these applications, such as single photon sources \cite{Kress2015} and SERS \cite{Wong2017}, the efficiency of the total emission into radiative channels (also sometimes termed the single-photon $\beta$-factor) is also very important.

As discussed earlier, the overall Raman enhancement of a single molecule as captured by the SMEF is attributed to both the $\textbf{E}$-field enhancement of the excitation and the LDOS enhancement for the Raman scattered light.  Here, we define the generalized Purcell factor that quantifies the enhancement of the LDOS for an emitter oriented in the $\textbf{n}$-direction in the vicinity of the WG, which is given by
\begin{equation}
\label{eq:Purcell_factor}
F_\textnormal{n}({\bf r}_m,\omega) = 1 + \frac{\textnormal{Im}\{\textbf{n}\cdot\textbf{G}_\textnormal{WG}(\textbf{r}_m,\textbf{r}_m;\omega)\cdot\textbf{n}\}}{\textnormal{Im}\{\textbf{n}\cdot\textbf{G}_{\rm B}(\textbf{r}_m,\textbf{r}_m;\omega)\cdot\textbf{n}\}},
\end{equation}
where we include a factor of 1 for molecules in free space \cite{Ge2014}. It must be noted that only a portion of the Raman scattered light is coupled to the WG mode, whereas the remaining portion is scattered into free space in the out-of-plane direction [Fig.\,\ref{fig:exp_config}(a)]. To quantify the efficiency with which the WG enhanced Raman scattering actually couples to the WG mode (in light of the fact that typically the Raman signal is collected via the WG mode), the WG $\beta$-factor can be defined as
\begin{equation}
\label{eq:beta_factor}
\beta({\bf r}_m,\omega) = \frac{\textnormal{Im}\{\textbf{n}\cdot\textbf{G}_\textnormal{WG}(\textbf{r}_m,\textbf{r}_m;\omega)\cdot\textbf{n}\}}{\textnormal{Im}\{\textbf{n}\cdot\textbf{G}(\textbf{r}_m,\textbf{r}_m;\omega)\cdot\textbf{n}\}}.
\end{equation}
This definition of the $\beta$-factor for WGs is consistent with previous work and specifically for plasmonic WGs \cite{Kress2015,Burmudez2015,Martin-Cano2015,Perera2013,Barthes2013,Chen2010,Chen2010_2,Jun2008}.

\subsection{Raman Enhancement of an Ensemble of Molecules}
\label{subsec:EF}

Although the SMEF assesses the Raman enhancement performance of a single molecule at a particular location in the vicinity of a WG, the more common scenario is that an ensemble of molecules is excited along the length of the WG and the accumulated Raman scattered power from them is measured. By utilizing the increased light-matter interaction volume via propagation of both excitation and Raman scattered light along the WG, the accumulated Raman scattered power can be much higher than that of the reference case when an analyte liquid in a cuvette is directly excited with a free space Gaussian beam. 

First, we define the spatially \textit{averaged} Raman enhancement factor (AEF), which is the average SMEF experienced by all molecules in the vicinity of the WG, since the Raman scattered power $P^\textnormal{SM}$ from a molecule is dependent on its cross-sectional position $\boldsymbol{\rho}_m$ as well as along the length of the WG $z_m$. As such,
\begin{align}
\label{eq:AEF}
\textnormal{AEF} &= \frac{\int_V \textnormal{SMEF}(\mathbf{r}_m)dV}{V} \nonumber \\  
&= \frac{\int_{z_m=0}^{L}\int_{\mathcal{A}_m}P^\textnormal{SM}(\mathbf{r}_m)d\boldsymbol{\rho}_mdz_m}{P^\textnormal{SM}_0 \mathcal{A}_m L},
\end{align}
where $\mathcal{A}_m$ is the cross-sectional area in the vicinity of the WG where light-matter interaction takes place, $L$ is the length of the WG, and $V = \mathcal{A}_m L$ is the volume over which spatial averaging is performed. For the slot WGs studied in this work, $\mathcal{A}_m$ is taken as the cross-sectional area of the hollow gap region. Note that the AEF is dependent on the excitation and Raman signal collection configurations [Fig.\,\ref{fig:exp_config}(a)] as is for SMEF. For each case, the corresponding formulation for $P^\textnormal{SM}$ [Eq.\,\eqref{eq:SERS_power}] as given in Subsection \ref{subsec:SMEF} is employed. The details of the calculation of AEF in Eq.\,\eqref{eq:AEF} and its derivation are given in Appendix \ref{Appendix:accum_enhancement}. By exploiting Eq.\,\eqref{eq:AEF}, the AEF in the forward, backward, and out-of-plane scattered directions can be calculated via simple analytical expressions derived in Appendix \ref{Appendix:accum_enhancement}.

While AEF only captures the SERS-type enhancement offered by the WG, the use of a WG structure also modifies the effective light-matter interaction volume compared to the interrogation volume of the free space Gaussian beam. Here, we define a figure-of-merit that captures the Raman enhancement of an ensemble of molecules by also taking into account the enhanced or reduced light-matter interaction volume offered by the WG, which we term the volume Raman enhancement factor (VEF), and it is given by
\begin{equation}
\label{eq:VEF}
\textnormal{VEF} = \textnormal{AEF}\frac{V}{V_0} = \textnormal{AEF}\frac{\mathcal{A}_m L}{\mathcal{A}_0 b}.
\end{equation}
This experimentally relevant enhancement factor is essentially AEF scaled by the ratio of the light-matter interaction volumes between the cases of employing the finite-size WG and using a Gaussian beam. In Eq.\,\eqref{eq:VEF}, $V_0 = \mathcal{A}_0 b$ is the estimated light-matter interaction volume near the Gaussian beam waist region, where $\mathcal{A}_0$ is the beam waist area and $b$ is the depth of focus. The waist area is given by $\mathcal{A}_0 = \pi(D_0/2)^2$, where the waist diameter is $D_0 = 1.22\lambda_0/\textnormal{NA}$, in which $\lambda_0$ is the free space wavelength and NA is the numerical aperture of the objective lens that focuses the Gaussian beam; the depth of focus is $b = 2\pi D_0^2n_B/\lambda_0$. Note that the Raman scattered power from each molecule within the Gaussian beam is essentially considered to be equivalent to the free space Raman scattered power $P_0^\textnormal{SM}$. It is also assumed that $P_0^\textnormal{SM}$ is the same for each molecule within the Gaussian beam waist region in which light-matter interaction occurs, and its calculation is detailed in Subsection \ref{subsec:SMEF}.

In our calculations, the effects of dipole-dipole interactions between molecules on Raman scattering is ignored, as it modifies the SERS-type Raman enhancement by a negligible amount for molecule separation distances of over two times the molecular diameter \cite{Chew1983}, due to the rapid decrease of dipole-dipole interaction with distance. For example, the diameter of a water molecule is $\sim$0.2 nm and the average inter-molecular spacing of its liquid phase is $\sim$0.4 nm, which means dipole-dipole interactions in liquids and solutions generally would not influence Raman enhancement of the sample, although it does have appreciable effects for specialized conditions involving very large and complex molecules such as polymers \cite{Kotula2017,Milani2008,Shimada2014}. Although it is possible for a Raman scattered photon to initiate another Raman scattering event in a different molecule, which constitutes the effect of stimulated Raman scattering given that the scattered intensity is comparable to that of the excitation, the required threshold pump intensity is on the order of several to hundreds of GW/$\textnormal{cm}^2$\,\,\cite{Gazengel1979,Gorelik2015}. Different strategies have been demonstrated to lower the threshold intensity for stimulated Raman scattering in liquids \cite{Gorelik2013,Gorelik2015}; however, the reduced values are still much higher than what are typically employed in Raman spectroscopy \cite{Mak2013}. As such, this effect does not have to be taken into account in our calculations of Raman enhancement.  

% =============================================================================================
% 

\section{Results}
\label{sec:Results}

\subsection{Plasmonic Slot Waveguide Designs: Purcell and Beta Factors, and Waveguide Mode Properties}
\label{sec:WG_mode_properties}

\begin{figure*}
	\centering
	\includegraphics[width=\textwidth]{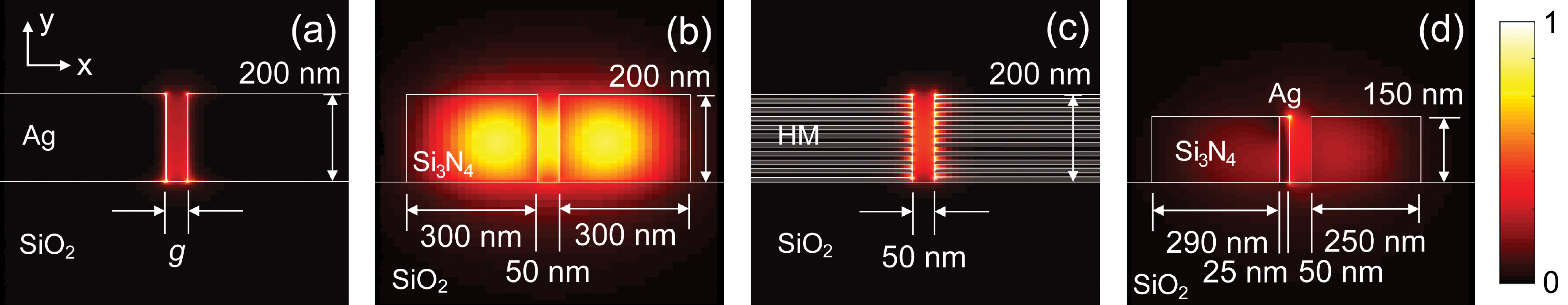}
	\caption{The $x$-$y$ cross-section and EM energy density distribution ${\cal w}(\boldsymbol{\rho}$) [J/$\textnormal{m}^3$] (see Appendix\,\ref{Appendix:energy_velocity} for derivation) of (a) Ag plasmonic slot WG (PSW) with gap width $g$, (b) $\textnormal{Si}_3\textnormal{N}_4$ dielectric slot WG (DSW), (c) Ag-$\textnormal{Si}_3\textnormal{N}_4$ HM plasmonic slot WG (HMSW) with $f_m$ = 0.5 (Ag metal layers are shown as gray), and (d) Ag-$\textnormal{Si}_3\textnormal{N}_4$ hybrid plasmonic slot WG (HPSW). The color bar is in linear scale (normalized units), and each plot is individually normalized.}
	\label{fig:wg_structures}
\end{figure*}

Amongst the different common types of metal based plasmonic WGs, the metal-insulator-metal structure offers the highest EM field confinement due to the presence of metal cladding layers on both sides surrounding the dielectric core \cite{Maier2007,Dionne2006}, which can be made arbitrarily thin. One manifestation of the metal-insulator-metal structure is the plasmonic slot WG, which consists of a thin metal film on top of a dielectric substrate that is subsequently etched to form a narrow slot tens to hundreds of nanometers wide \cite{Dionne2006,Atwater2007,Gramotnev2010}. The slot (or gap) region of the plasmonic slot WG structure where high EM field concentration resides is also amenable for filling by analyte molecules of interest, which is suitable for SE and Raman enhancement applications. The plasmonic slot WG structure that we employ here is shown in Fig.\,\ref{fig:wg_structures}(a), using a 200 nm thick silver (Ag) film on top of silica ($\textnormal{SiO}_2$) substrate with a gap width $g$ = 50 nm. 

We also investigate a few other selected types of slot WG structures. One example is an integrated dielectric WG, which has already been demonstrated experimentally for use in Raman spectroscopy \cite{Dhakal2014,Dhakal2015,Dhakal2016,Evans2016,Holmstrom2016}; thus one of the WG structures we study here is the silicon nitride ($\textnormal{Si}_3\textnormal{N}_4$) dielectric slot WG as shown in Fig.\,\ref{fig:wg_structures}(b), because it would be beneficial to compare our proposed plasmonic type WGs (which to our knowledge has not previously been applied to Raman spectroscopy) to the performance of dielectric WGs in an on-chip setting. Another type of WG being investigated is the stratified hyperbolic material (HM) slot WG that is shown in Fig.\,\ref{fig:wg_structures}(c), for which Ag and $\textnormal{Si}_3\textnormal{N}_4$ are utilized in the alternating multilayer stack. Hyperbolic materials have been shown to exhibit ultra-high LDOS and thus large enhancement of SE rates of emitters located near the surface or embedded in the bulk \cite{Jacob2012,Noginov2010,Newman2013,Ferrari2014,Shalaginov2015}. More recently, nanoresonators constructed from HMs have also been investigated for both SE \cite{Axelrod2017,Slobozhanyuk2015} and Raman \cite{Wong2017} enhancement. Waveguides that utilize HMs have also been studied for different properties and applications \cite{Mironov2014,Babicheva2015,He2012,He2012_2,Ishii2014}, including for SE enhancement \cite{Roth2017}. Finally, we also consider the hybrid plasmonic slot WG [Fig.\,\ref{fig:wg_structures}(d)], in which the guided mode is a hybrid between a surface plasmon polariton and a total-internal-reflection mode; the structure investigated contains Ag as the metal component and $\textnormal{Si}_3\textnormal{N}_4$ for the dielectric regions. The hybrid plasmonic WGs exhibit improved propagation lengths compared to pure metal plasmonic WGs, while maintaining high mode confinement in the sub-wavelength scale \cite{Alam2010,Ma2014}. In our study, the background medium is aqueous, so that the refractive index of water, $n_\textnormal{B}$ = 1.33, is utilized. 

\begin{figure}
	\centering
	\includegraphics[width=\columnwidth]{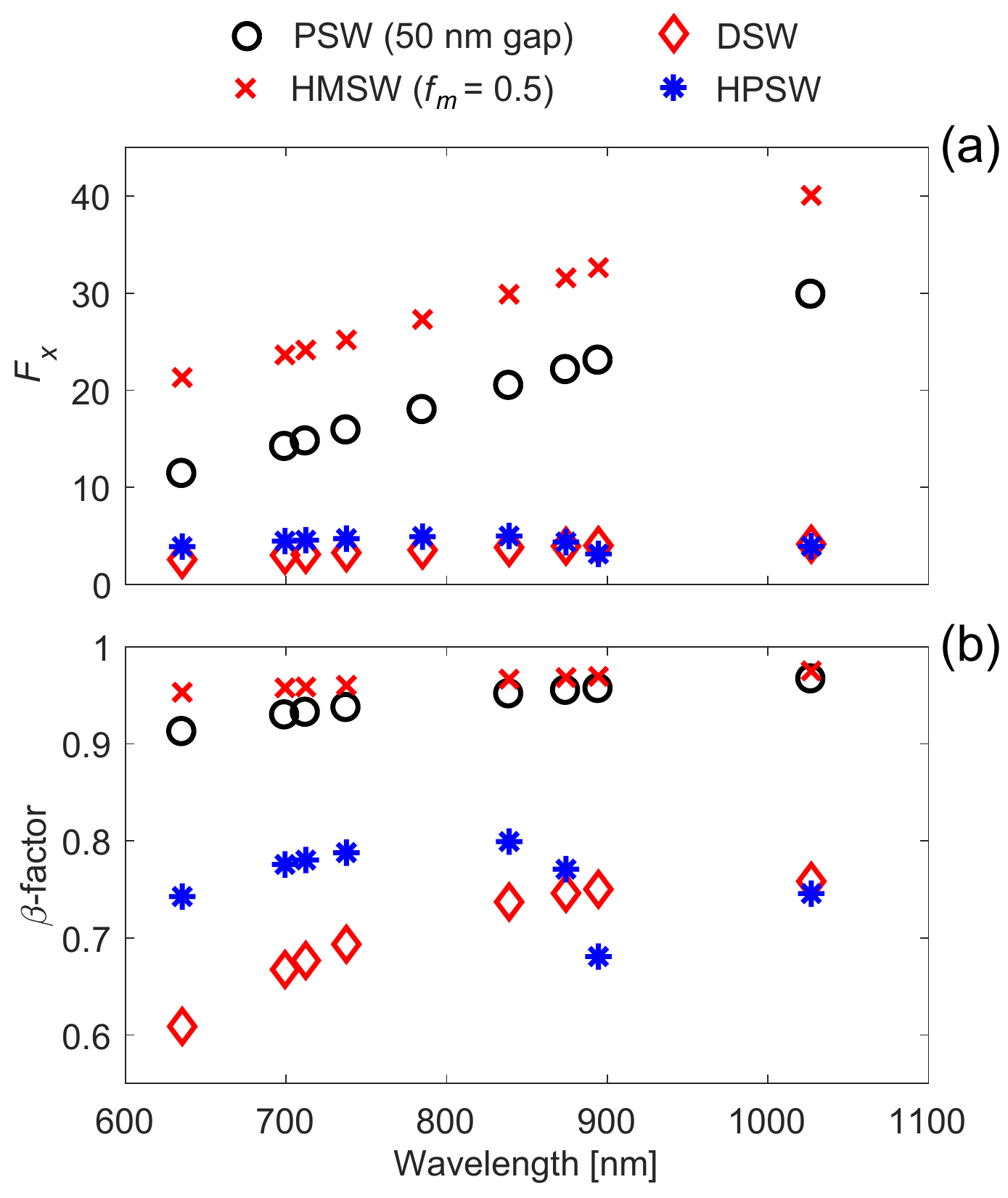}
	\caption{(a) Purcell factor for dipole oriented in the $x$-direction $F_x$ and (b) WG mode $\beta$-factor (Eq.\,\eqref{eq:beta_factor}) as a function of wavelength for different types of plasmonic slot WGs (PSWs) as shown in Fig.\,\ref{fig:wg_structures}. Black circle: Ag PSW with gap width $g$ = 50 nm; Red diamond: $\textnormal{Si}_3\textnormal{N}_4$ dielectric slot WG (DSW); Red cross: Ag-$\textnormal{Si}_3\textnormal{N}_4$ HM plasmonic slot WG (HMSW) with $f_m$ = 0.5; Blue asterisk: Ag-$\textnormal{Si}_3\textnormal{N}_4$ hybrid plasmonic slot WG (HPSW). Each marker indicates either the pump wavelength or a specific Stokes or anti-Stokes wavelength corresponding to one of the Raman modes as described in Section\,\ref{sec:WG_mode_properties}.}
	\label{fig:Fp_beta}
\end{figure}

\begin{figure*}
	\centering
	\includegraphics[width=\textwidth]{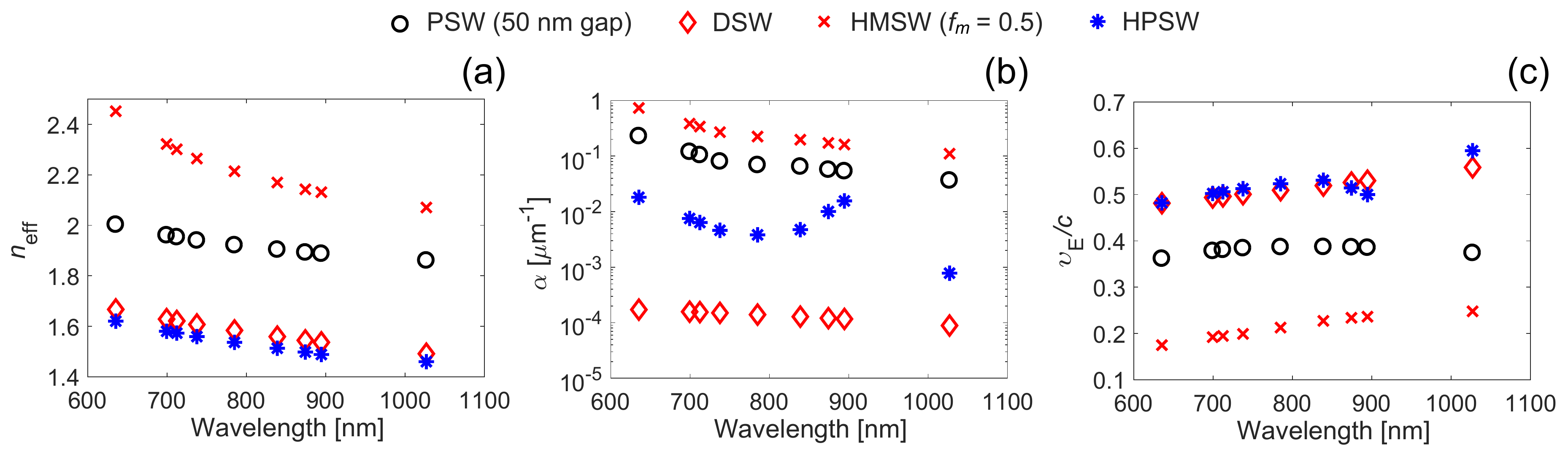}
	\caption{(a) Effective index $n_\textnormal{eff}$, (b) modal loss $\alpha$, and (c) normalized energy velocity $\upsilon_\textnormal{E}$/c as a function of wavelength for different types of plasmonic slot WGs (PSWs) as shown in Fig.\,\ref{fig:wg_structures}. Black circle: Ag PSW with gap width $g$ = 50 nm; Red diamond: $\textnormal{Si}_3\textnormal{N}_4$ dielectric slot WG (DSW); Red cross: Ag-$\textnormal{Si}_3\textnormal{N}_4$ HM plasmonic slot WG (HMSW) with $f_m$ = 0.5; Blue asterisk: Ag-$\textnormal{Si}_3\textnormal{N}_4$ hybrid plasmonic slot WG (HPSW). Each marker indicates either the pump wavelength or a specific Stokes or anti-Stokes wavelength corresponding to one of the Raman modes as described in Section\,\ref{sec:WG_mode_properties}.}
	\label{fig:WGs_dispersion}
\end{figure*}

As observed in Fig.\,\ref{fig:Fp_beta}(a), the Purcell factor for a dipole that is oriented in the $x$-direction $F_x$ (located at the center of the WG gap region) can be increased significantly by employing the Ag plasmonic slot WG compared to using a $\textnormal{Si}_3\textnormal{N}_4$ dielectric slot WG; e.g., at the wavelength $\lambda_0$ = 785 nm, $F_x$ improves from 3.5 to 18.0, and the enhancement exists over a broad bandwidth. This is first attributed to the much lower energy velocity $\upsilon_\textnormal{E}$ of the plasmonic slot WG mode, e.g., with $\upsilon_\textnormal{E}/c$ = 0.39 compared to $\upsilon_\textnormal{E}/c$ = 0.51 for the $\textnormal{Si}_3\textnormal{N}_4$ slot WG at $\lambda_0$ = 785 nm [Fig.\,\ref{fig:WGs_dispersion}(c)]. Secondly, the EM energy density ${\mathcal w}(\boldsymbol{\rho})$, and thus the $\textbf{E}$-field amplitude $\textbf{E}$, is predominantly concentrated within the gap region of the Ag plasmonic slot WG [Fig.\,\ref{fig:wg_structures}(a)], in comparison to the $\textnormal{Si}_3\textnormal{N}_4$ slot WG in which ${\cal w}(\boldsymbol{\rho})$ also resides prominently inside the dielectric claddings [Fig.\,\ref{fig:wg_structures}(b)]. This leads to a much higher normalized $\textbf{E}$-field $\textbf{e}_{k_\omega}(\boldsymbol{\rho})$ within the gap region, as shown by Eq.\,\eqref{eq:E_field_normalization_in_Appendix} in Appendix \ref{Appendix:energy_velocity}. The decrease in $\upsilon_\textnormal{E}$ and increase in $\textbf{e}_{k_\omega}(\boldsymbol{\rho})$ within the gap region by using a plasmonic slot WG together enhances $\textbf{G}_\textnormal{WG}$ [Eq.\,\eqref{eq:Green_function}] compared to using a dielectric slot WG, and thus improves the Purcell factor $F_x$ [Eq.\,\eqref{eq:Purcell_factor}]. The benefits of using a plasmonic slot WG extends to also improving the $\beta$-factor, as can be observed from Fig.\,\ref{fig:Fp_beta}(b); at $\lambda_0$ = 839 nm, $\beta$ increases from 0.74 for the $\textnormal{Si}_3\textnormal{N}_4$ slot WG to 0.95 for the Ag plasmonic slot WG. Indeed, there is benefit in moving to the use of metal based plasmonic slot WGs for applications requiring LDOS and SE enhancement, instead of employing dielectric type WGs. However, the drawback of using plasmonic WGs is that the modal loss ($\alpha = 2\kappa$) is adversely increased by over an order of magnitude; as seen in Fig.\,\ref{fig:WGs_dispersion}(b), the loss increases from $\alpha$ = 1.40 $\times$ $10^{-4}$ $\mu \textnormal{m}^{-1}$ for the $\textnormal{Si}_3\textnormal{N}_4$ slot WG to $\alpha$ = 6.91 $\times$ $10^{-2}$ $\mu \textnormal{m}^{-1}$ for the Ag plasmonic slot WG at $\lambda_0$ = 785 nm. Such an increase in modal loss has significant implications for many applications; we will show subsequently in this paper how it affects the design of WGs for Raman enhancement.

Next, we investigate the slot WG constructed from a stratified metal-dielectric stack that forms a HM, in which the thickness of each metal or dielectric layer is chosen to be 10 nm [Fig.\,\ref{fig:wg_structures}(c)], and thus the metal filling fraction (volume fraction of metallic material) is $f_m$ = 0.5. It can be observed from Fig.\,\ref{fig:Fp_beta}(a) that the Purcell factor is further increased compared to that of the Ag plasmonic slot WG, to $F_x$ = 27.3 at $\lambda_0$ = 785 nm. The $\beta$-factor is also slightly improved [Fig.\,\ref{fig:Fp_beta}(b)], such that it becomes $\beta$ = 0.97 at $\lambda_0$ = 839 nm. The HM slot WG is able to attain higher Purcell and $\beta$-factors, because its guided mode has a much lower energy velocity, such that $\upsilon_\textnormal{E}/c$ = 0.21 at $\lambda_0$ = 785 nm [Fig.\,\ref{fig:WGs_dispersion}(c)]. Again, this comes at the expense of higher modal loss; $\alpha$ = 0.224 $\mu \textnormal{m}^{-1}$ at $\lambda_0$ = 785 nm.

To help alleviate the high modal loss of the plasmonic slot WG while maintaining high $\textbf{E}$-field concentration within the gap region, the hybrid plasmonic slot WG [Fig.\,\ref{fig:wg_structures}(d)] is studied. The flexibility of this WG structure is that it can be optimized for either minimum modal loss or maximum fraction of $\textbf{E}$-field intensity within the gap region, simply by tuning the width of the left $\textnormal{Si}_3\textnormal{N}_4$ cladding while keeping all other WG dimensions constant. Hybrid plasmonic WGs with very similar structures as we study here have been demonstrated for several applications \cite{Ma2014,Su2017,Wong2018}; in each case the optimal device was also obtained by tuning the thicknesses of the constituent layers. The details of the optimization of the hybrid plasmonic slot WG are shown in Fig.\,\ref{fig:Hybrid_wg_sweep} in Appendix \ref{Appendix:additional_figures}, where the optimal left $\textnormal{Si}_3\textnormal{N}_4$ cladding width for minimal modal loss is 90 nm, and the optimal width for maximum $\textbf{E}$-field intensity within the gap region is 290 nm. Here, we choose to present the results of the hybrid plasmonic slot WG optimized for maximal $\textbf{E}$-field intensity in the gap, because it exhibits better Raman enhancement performance. It is seen in Fig.\,\ref{fig:WGs_dispersion}(b) that although the modal loss of the hybrid plasmonic slot WG is indeed much lower than that of the Ag plasmonic WG ($\alpha$ = 3.82 $\times$ $10^{-3}$ $\mu \textnormal{m}^{-1}$ at $\lambda_0$ = 785 nm), it is still significantly higher than for the $\textnormal{Si}_3\textnormal{N}_4$ slot WG. However, the Purcell factor is quite similar to that attained by the $\textnormal{Si}_3\textnormal{N}_4$ slot WG across all wavelengths from 600 to 1100 nm [Fig.\,\ref{fig:Fp_beta}(a)], although the WG mode $\beta$-factor is appreciably improved for wavelengths $\lambda_0$ $\lesssim$ 880 nm [Fig.\,\ref{fig:Fp_beta}(b)]. Overall, the hybrid plasmonic slot WG provides very little benefits compared to a much simpler purely dielectric WG structure for SE enhancement applications. 

The ability of plasmonic slot WGs (either using Ag or HM) to drastically improve the Purcell factor compared to the dielectric slot WG is due to the much reduced $\upsilon_\textnormal{E}$ as well as increased concentration of EM fields within the gap region where the dipole sits. This is enabled by a completely different waveguiding mechanism, which is via surface plasmon polaritons, as opposed to total-internal-reflection in dielectric WGs. For a given plasmonic WG structure, as the frequency of operation $\omega$ increases (or wavelength $\lambda_0$ decreases) to approach the surface plasmon frequency $\omega_\textnormal{sp}$, the slope of the $\omega$-$k$ dispersion decreases, which leads to a decrease of not only the energy velocity $\upsilon_\textnormal{E}$ [Fig.\,\ref{fig:WGs_dispersion}(c)] but also a decrease in the phase velocity \cite{Maier2007,Dionne2006}; both of which can be much lower than using dielectric WGs. A decrease in phase velocity is equivalent to an increase in the effective index $n_\textnormal{eff} = k/k_0$, where $k_0 = 2\pi/\lambda_0$ is the free space wave vector. It is seen in Fig.\,\ref{fig:WGs_dispersion}(a) that $n_\textnormal{eff}$ is much higher for the Ag plasmonic slot WG compared to the $\textnormal{Si}_3\textnormal{N}_4$ slot WG, and that it also increases as $\lambda_0$ decreases. The use of a stratified HM multilayer stack for the plasmonic slot WG [Fig.\,\ref{fig:wg_structures}(c)] serves to alter the effective permittivity of the material; it becomes anisotropic such that $\varepsilon_x = \varepsilon_z < 0$, but it is not equivalent to the permittivity in the $y$-direction ($\varepsilon_y > 0$) \cite{Ferrari2015,Poddubny2013,Shekhar2014}. Using HMs essentially reduces the metal filling fraction $f_m$ and thus it is ``less metallic" compared to pure metal (in the $x$-direction), which is responsible for reducing $\omega_\textnormal{sp}$ (or increasing $\lambda_\textnormal{sp}$). The effect is that at a given operating wavelength $\lambda_0$, it is closer to $\lambda_\textnormal{sp}$ when HM is utilized, which means that $n_\textnormal{eff}$ is higher and $\upsilon_\textnormal{E}$ is lower compared to when Ag is utilized for the plasmonic slot WG, as shown in Figs.\,\ref{fig:WGs_dispersion}a and \ref{fig:WGs_dispersion}c, respectively.

\subsection{Raman Enhancement in Plasmonic Slot Waveguide}
\label{sec:Ag_PSW}

\begin{figure}
	\centering
	\includegraphics[width=\columnwidth]{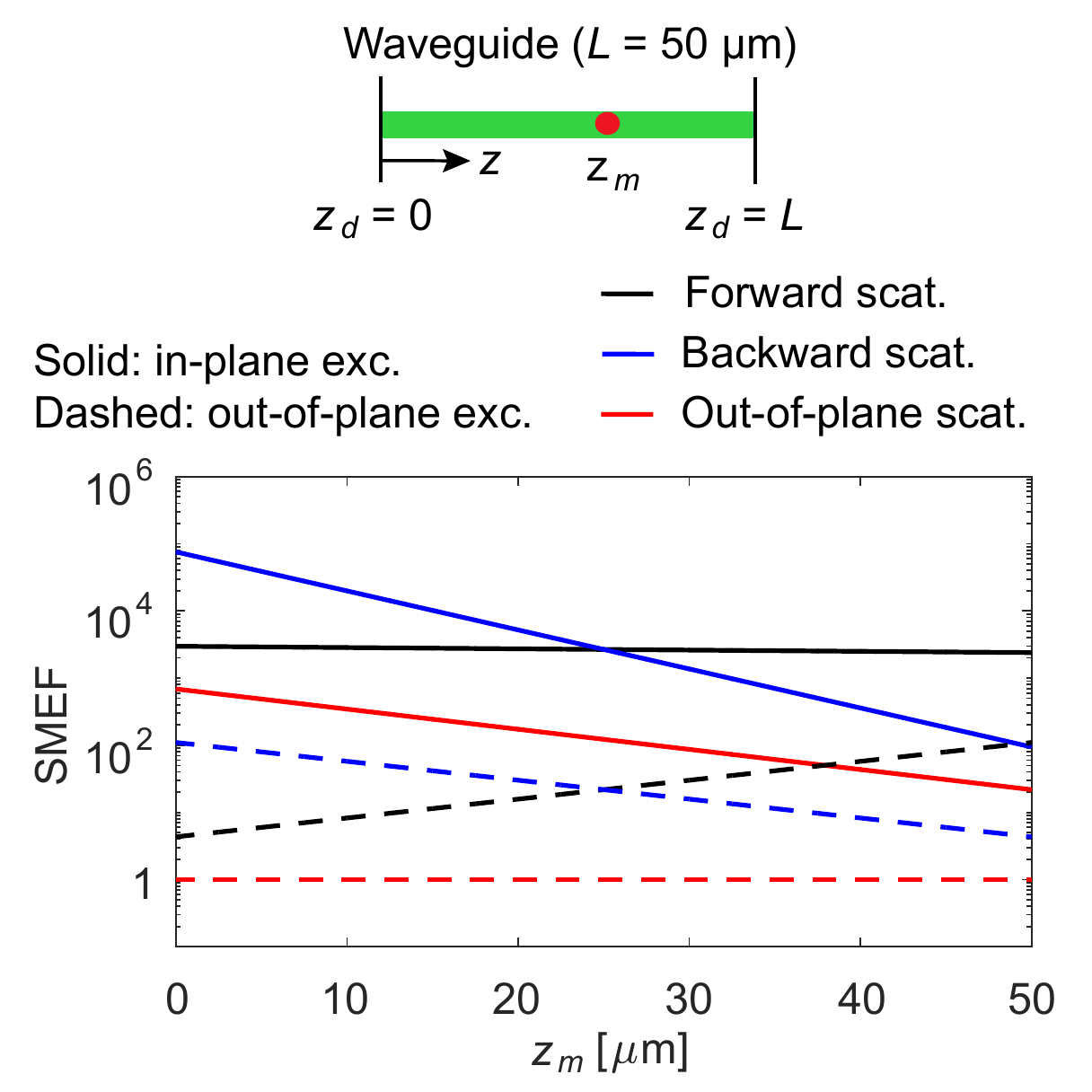}
	\caption{Single-molecule enhancement factor (SMEF) of a molecule at the center of the gap region in terms of $x$-$y$ position of a Ag plasmonic slot WG, with gap width $g$ = 50 nm [Fig.\,\ref{fig:wg_structures}(a)] and length $L$ = 50 $\mu\textnormal{m}$ as a function of its location $z_m$ along the WG length (for Stokes line of Raman mode with $\nu_{m,1}$). Solid: excitation coupled in-plane via the WG mode; dashed: excitation from out-of-plane. Black: forward scattered ($z_d$ = $L$); blue: backward scattered ($z_d$ = 0); red: out-of-plane scattered.}
	\label{fig:SMEF}
\end{figure}

\begin{figure}
	\centering
	\includegraphics[width=\columnwidth]{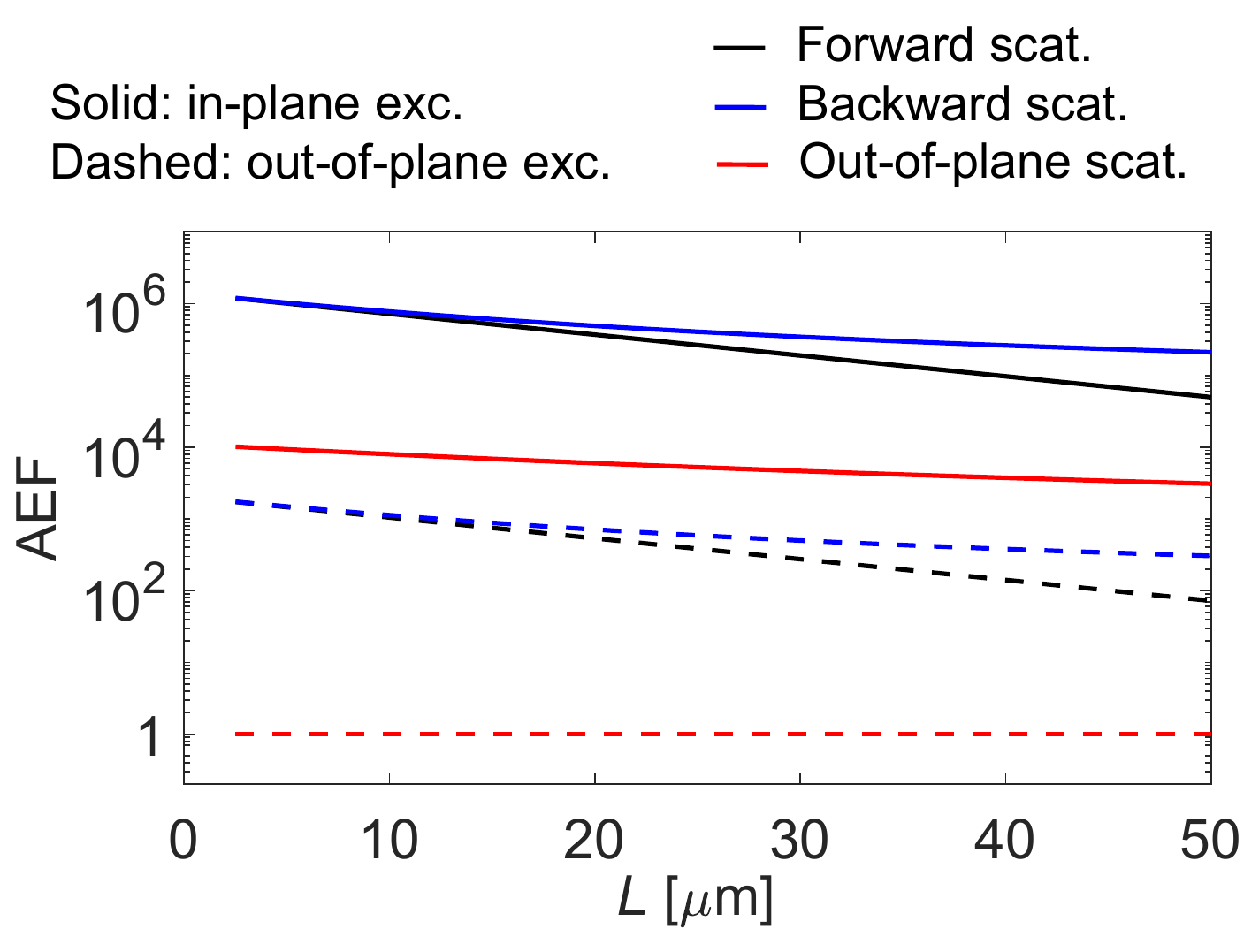}
	\caption{(a) Spatially averaged enhancement factor (AEF) for molecules within the gap region of an Ag plasmonic slot WG with gap width $g$ = 50 nm as a function of WG length $L$ for different excitation and Raman signal collection configurations (for Stokes line of Raman mode with $\nu_{m,1}$). Solid: excitation coupled in-plane via the WG mode; dashed: excitation from out-of-plane. Black: forward scattered; blue: backward scattered; red: out-of-plane scattered.}
	\label{fig:Avg_EF}
\end{figure}

\begin{figure}
	\centering
	\includegraphics[width=\columnwidth]{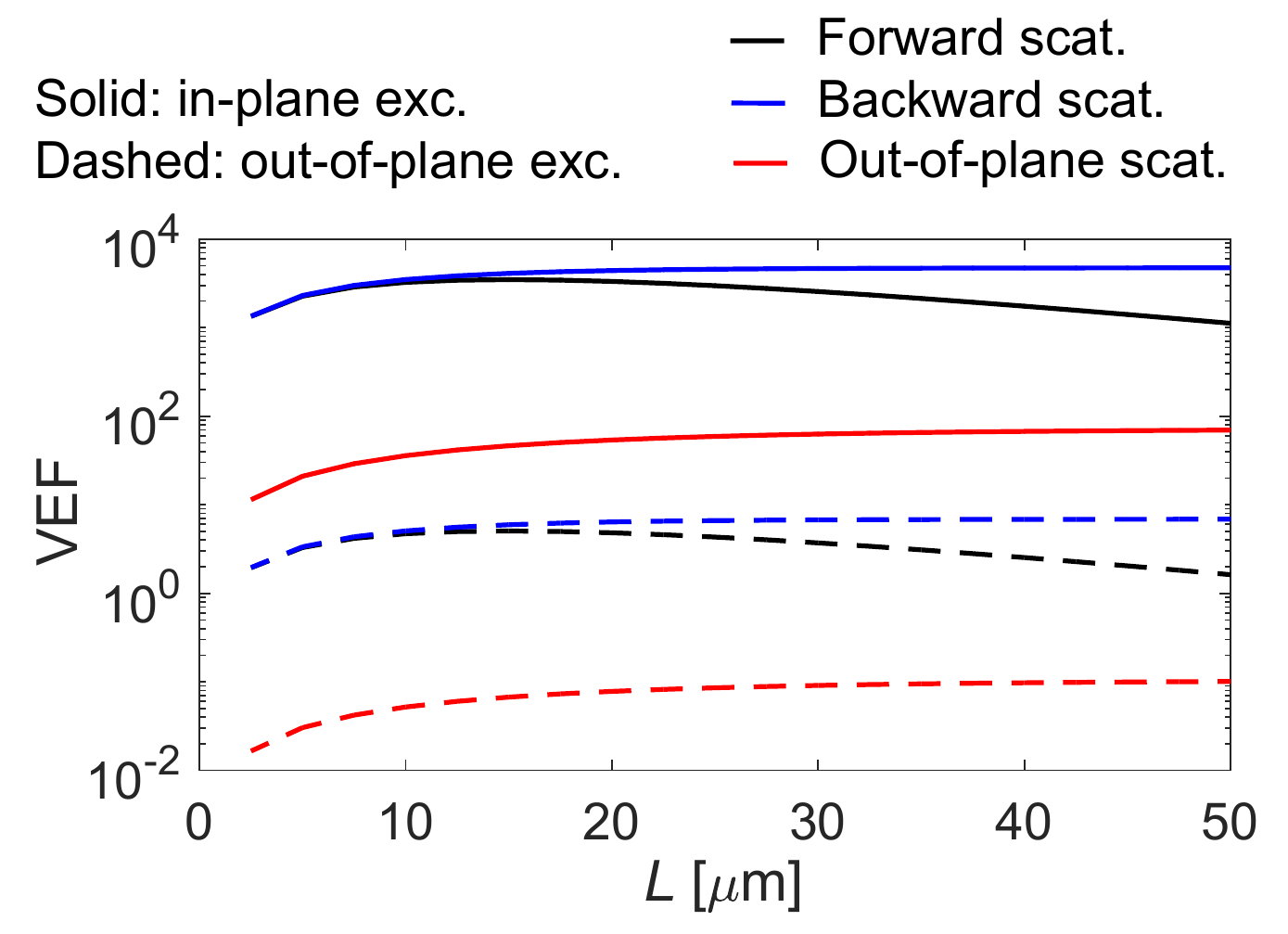}
	\caption{(a) Volume enhancement factor (VEF) for molecules within the gap region of an Ag plasmonic slot WG with gap width $g$ = 50 nm as a function of WG length $L$ for different excitation and Raman signal collection configurations (for Stokes line of Raman mode with $\nu_{m,1}$). The results here are based on comparing to a reference Gaussian beam focused using an objective lens with NA = 0.75 that has a waist diameter $D_0 \approx$ 1.28 $\mu\textnormal{m}$ and depth of focus $b \approx$ 17.4 $\mu\textnormal{m}$. Note that changing the Gaussian beam parameters will simply scale the VEF. Solid: excitation coupled in-plane via the WG mode; dashed: excitation from out-of-plane. Black: forward scattered; blue: backward scattered; red: out-of-plane scattered.}
	\label{fig:Accum_EF}
\end{figure}

To better highlight the main characteristics of utilizing WGs for Raman enhancement, we first focus on only the Ag plasmonic slot WG with gap width $g$ = 50 nm, as shown in Fig.\,\ref{fig:wg_structures}(a). Here, we model the Raman active molecule Rhodamine 6G (R6G), and present results for the Stokes line of one of its Raman modes with shift $\nu_m$ = 819 $\textnormal{cm}^{-1}$ ($\omega_m$ = 98.7 meV, $\textnormal{RA} \triangleq R_{nn}^2$ = 6.2 $\mathring{\textnormal{A}}^4 \textnormal{amu}^{-1}$ = 3.73 $\times 10^{-13}$ $\textnormal{m}^4\textnormal{kg}^{-1}$). The decay rate of the Raman mode $\gamma_m$ is taken to be 1.6 meV. An excitation wavelength of 785 nm and an intensity of 10 $\textnormal{mW}/\mu\textnormal{m}^2$ are utilized for the calculations, which are representative parameters in typical Raman spectroscopy systems \cite{Dhakal2016,Mak2013}. In Appendix \ref{Appendix:EF_diff_Raman_modes}, the results of Raman enhancement by using the Ag plasmonic slot WG with $g$ = 50 nm for several other Raman modes of R6G are presented.

First, we consider the scenario in which a single R6G molecule is located at the center of the gap region in the $x$-$y$ plane, which is defined as the position that is 100 nm above the $\textnormal{SiO}_2$ substrate and half way between the two adjacent vertical walls of the slot region (Fig.\,\ref{fig:wg_structures}), and its location $z_m$ is varied along the length of the WG in the $z$-direction. The WG considered has a length $L$ = 50 $\mu\textnormal{m}$ (Fig.\,\ref{fig:SMEF}). As the molecule is translated away from the input facet at $z = 0$ to increase $z_m$, the $\textbf{E}$-field enhancement factor $|\eta(z_m)|$ decreases because the pump power decays as a function of distance from the input facet to the molecule location $z_m$ due to WG modal loss at the pump wavelength, and this means $|\eta(z_m)| \propto e^{-2\kappa z_m}$ [Eq.\,\eqref{eq:field_enhancement}]; $|\eta(z_m)|$ is maximized when $z_m$ = 0. Simultaneously, the distances between the molecule to both the input and end facets are changed, which also influences the power of the Raman scattered light $P^\textnormal{SM}$ at the detection location $z_d$ for each of the cases of forward ($z_d = L$) and backward ($z_d$ = 0) scattered configurations (Fig.\,\ref{fig:SMEF}); this is due to the modal loss at the Raman scattered wavelength. In the forward scattering case, the propagation loss of the Raman scattered light leads to transmission to the end facet at $z_m = L$ to be proportional to $e^{-2\kappa(L - z_m)}$; the overall detected Raman scattered power $P^\textnormal{SM}$ and thus SMEF is proportional to $e^{-2\kappa z_m}e^{-2\kappa(L - z_m)} = e^{-2\kappa L}$, which means that it is independent of the molecule location $z_m$ (solid black line in Fig.\,\ref{fig:SMEF}). For backward scattering, the transmission back to the input facet at $z_m = 0$ is proportional to $e^{-2\kappa z_m}$, so that $P^\textnormal{SM}$ and thus SMEF is proportional to $e^{-2\kappa z_m}e^{-2\kappa z_m} = e^{-4\kappa z_m}$, which decays linearly on a log-scale as shown by the solid blue line in Fig.\,\ref{fig:SMEF}. Note that the modal EM field distribution at the 2-D cross-section is invariant along the $z$-direction, but only the total field amplitude or mode power decays. Finally, for the out-of-plane Raman scattered signal, the detected signal is not dependent on the modal loss at the Raman scattered wavelength, so $P^\textnormal{SM}$ and thus SMEF is proportional to $e^{-2\kappa z_m}$, which leads to a linear decay with a smaller slope in Fig.\,\ref{fig:SMEF} (solid red line). Similar arguments can be made for the out-of-plane excitation configuration, in which the detected Raman signal would only depend on the modal loss at the Raman scattered wavelength, but not on that at the excitation wavelength (Fig.\,\ref{fig:SMEF}).

For the Ag plasmonic slot WG, the SMEF attained by in-plane excitation via the WG mode is in general much higher than that by out-of-plane excitation from the top (by at least one-order of magnitude), for all molecule locations $z_m$ (Fig.\,\ref{fig:SMEF}). This observation emphasizes the significance of utilizing the plasmonic slot WG mode for enhancing the excitation field in order to improve the overall Raman enhancement. By collecting the Raman signal that is either forward or backward scattered via the WG mode, the attainable SMEF is always higher than by collecting out-of-plane, regardless of the molecule location $z_m$, which is due to the Purcell enhancement offered by the plasmonic slot WG. For the case of excitation via the WG mode and collection by out-of-plane scattering, the achievable SMEF is significantly higher than the case of out-of-plane pump coupling and collection via the WG mode (except for when $z_m$ $>$ 37.5 $\mu\textnormal{m}$ in the forward scattering case), which suggests that by using the plasmonic slot WG, the contribution to overall Raman enhancement by excitation field enhancement is actually stronger than by scattered field enhancement. This is in stark contrast to the use of nanostructured metal dimers as in traditional SERS, in which the contribution to Raman enhancement via excitation field enhancement is weaker than due to the enhancement of the scattered field \cite{Wong2017}. Finally, the red dashed line in Fig.\,\ref{fig:SMEF} indicates the case when both excitation and Raman signal collection are via the out-of-plane configuration, which is similar to the molecule situated in free space (and theoretically identical in our theory), so that each of the $\textbf{E}$-field enhancement $|\eta|$ and Purcell factor $F_x$ is assumed to be unity, and thus SMEF = 1 for all $z_m$, as expected. Overall, the maximum SMEF achievable using the Ag plasmonic slot WG is on the order of $10^4$ to $10^5$, which is approximately one-order of magnitude lower than the SMEF observed in metal dimers with similar gap widths \cite{Wong2017}.

The SMEF as defined thus far is based on a molecule situated at the center of the gap region of the Ag plasmonic slot WG. However, typically a bulk sample with an ensemble of molecules is measured in Raman spectroscopy, so the spatially averaged Raman enhancement factor (AEF) would be of importance. The AEF as a function of WG length $L$ for the different combinations of pump excitation and Raman scattered signal collection configurations are shown in Fig.\,\ref{fig:Avg_EF}. It is observed that in general, the AEF is much higher than the SMEF when comparing between the values attained using the same excitation and Raman collection configuration. For example, by employing a Ag plasmonic slot WG with $L$ = 50 $\mu m$ and considering the case of in-plane excitation and Raman signal collection in the backward scattered direction, the SMEF of a molecule at the center along the length ($z_m$ = 25 $\mu m$) is 2670 (Fig.\,\ref{fig:SMEF}); however, the AEF considering all molecules within the WG is $\sim$2.1 $\times 10^5$ (Fig.\,\ref{fig:Avg_EF}), which is close to two-orders of magnitude higher. The much higher AEF compared to SMEF is attributed to the lower $\textbf{E}$-field amplitude at the center of the Ag plasmonic slot WG gap region in comparison to at positions closer to the metal walls. The AEF by in-plane excitation of the WG mode is also higher than by coupling from free space, regardless of the Raman signal collection configuration.

In terms of Raman enhancement when measurement on a bulk sample with an ensemble of molecules is carried out, the volume enhancement factor (VEF) is defined that also takes into account any increase or decrease in light-matter interacton volume [Eq.\,\eqref{eq:VEF}]. The reference case that we consider is the Gaussian beam focused by an objective lens with NA = 0.75, which would have a waist diameter $D_0$ $\approx$ 1.28 $\mu\textnormal{m}$ and depth of focus $b$ $\approx$ 17.4 $\mu\textnormal{m}$. The VEF as a function of WG length $L$ for the different combinations of pump excitation and Raman scattered signal collection configurations are shown in Fig.\,\ref{fig:Accum_EF}. While there is an optimal $L$ for maximum VEF for the forward scattered configuration, the VEF actually plateaus as $L$ is increased for the backward scattered case. The trends in VEF comparing between different combinations of excitation and Raman scattering collection configurations are the same as for the AEF as shown in Fig.\,\ref{fig:SMEF}. In general, coupling the excitation beam into the WG mode achieves much higher VEF compared to excitation from out-of-plane, regardless of which configuration is used to collect the Raman scattered light. From Fig.\,\ref{fig:Accum_EF}, the red dashed line represents the case of out-of-plane configuration for both excitation and Raman signal collection, and VEF $<$ 1 for all WG lengths, which is attributed to the fact that the gap region of the Ag plasmonic slot WG has a volume that is much smaller than the Gaussian beam waist region. As such, without contributions from SERS-type enhancement, the overall Raman scattered signal for an ensemble of molecules is much lower than by direct Gaussian beam excitation. It is observed that the VEF (Fig.\,\ref{fig:Accum_EF}), for the chosen Gaussian beam reference, is in general much lower than the AEF (Fig.\,\ref{fig:Avg_EF}) using the Ag plasmonic slot WG, which indicates that in fact there is a lower interrogation volume by using the current design in comparison to using the Gaussian beam in free space. Nonetheless, the VEF quantifies the ability to provide Raman enhancement for an ensemble of molecules in a bulk sample, and it is shown that employing our proposed Ag plasmonic slot WG can lead to Raman enhancement factors that are close to four-orders of magnitude higher compared to by simple Gaussian beam excitation.

Taking the Stokes line of the Raman mode with shift $\omega_m$ = 98.7 meV at $\lambda_0$ = 839 nm, the maximum VEF of the forward scattered signal is 3480 using a Ag plasmonic slot WG with length $L$ = 15 $\mu\textnormal{m}$ (Fig.\,\ref{fig:Accum_EF}). In Ref. \citenum{Mak2013}, using a HC-PCF with the core filled with water, the intensity of the OH stretching mode at $\nu_m$ $\approx$ 3400 $\textnormal{cm}^{-1}$ ($\omega_m$ = 421.5 meV) is obtained from the Raman spectrum and compared to the reference case of measurement using a cover slide with water on top. The equivalent Raman enhancement factor by using a 4-cm long HC-PCF is EF $\approx$ 98. This means that our proposed Ag plasmonic slot WG can achieve $\sim$35$\times$ higher Raman enhancement in an on-chip integrated device that is over three-orders of magnitude smaller in length and more than an order of magnitude less in cross-section, compared to this experimental work.

% =============================================================================================

\subsection{Comparison of Raman Enhancement Between Different Waveguides}
\label{sec:Comparison_diff_wgs}

\begin{figure*}
	\centering
	\includegraphics[width=0.95\textwidth]{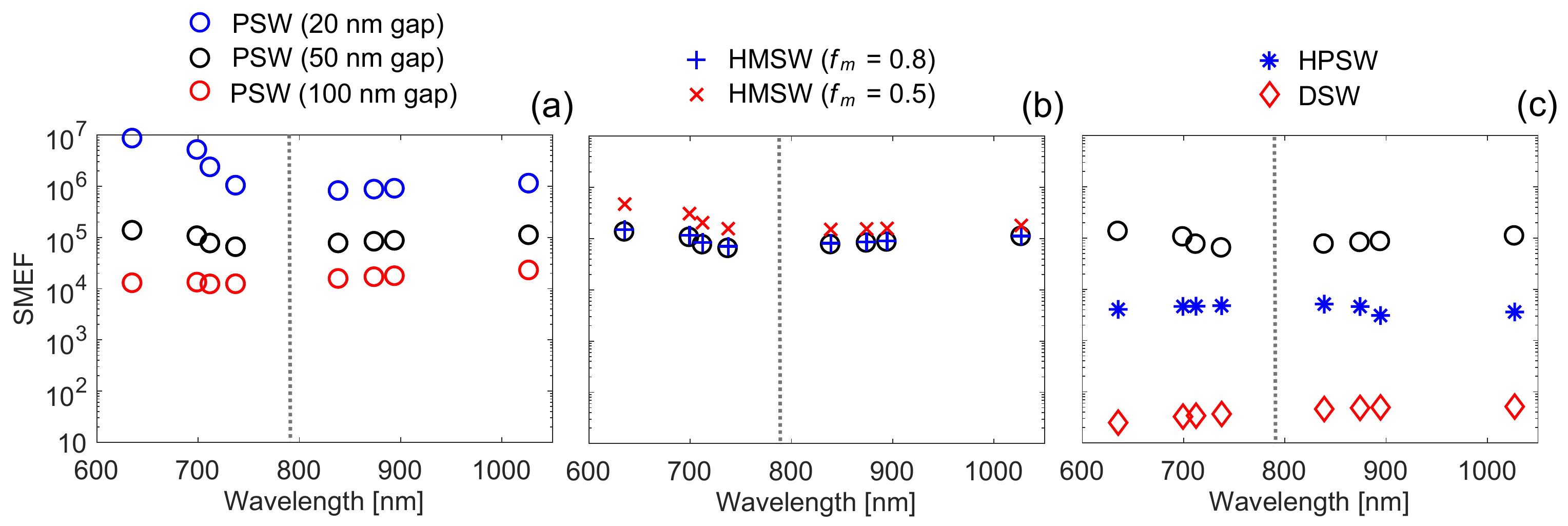}
	\caption{Single-molecule enhancement factor (SMEF) for when the molecule location $z_m$ = 0 and by in-plane excitation and collection of backward scattered signal (see Fig.\,\ref{fig:SMEF}), as a function of wavelength for the Ag plasmonic slot WG (PSW), with gap width $g$ = 50 nm, and other WGs obtained by (a) varying the gap width $g$: Ag PSW with gap width $g$ = 100 nm and $g$ = 20 nm, (b) varying the metal filling fraction $f_m$: Ag-$\textnormal{Si}_3\textnormal{N}_4$ HM slot WG (HMSW) with $f_m$ = 0.8 and $f_m$ = 0.5, and (c) changing the WG type: Ag-$\textnormal{Si}_3\textnormal{N}_4$ hybrid plasmonic slot WG (HPSW) and $\textnormal{Si}_3\textnormal{N}_4$ dielectric slot WG (DSW). Each marker indicates a specific Stokes or anti-Stokes wavelength corresponding to one of the Raman modes as described in Section\,\ref{sec:WG_mode_properties}. The vertical dotted line indicates the pump wavelength at $\lambda_0$ = 785 nm.}
	\label{fig:SMEF_vs_wavelength}
\end{figure*}

\begin{figure*}
	\centering
	\includegraphics[width=0.95\textwidth]{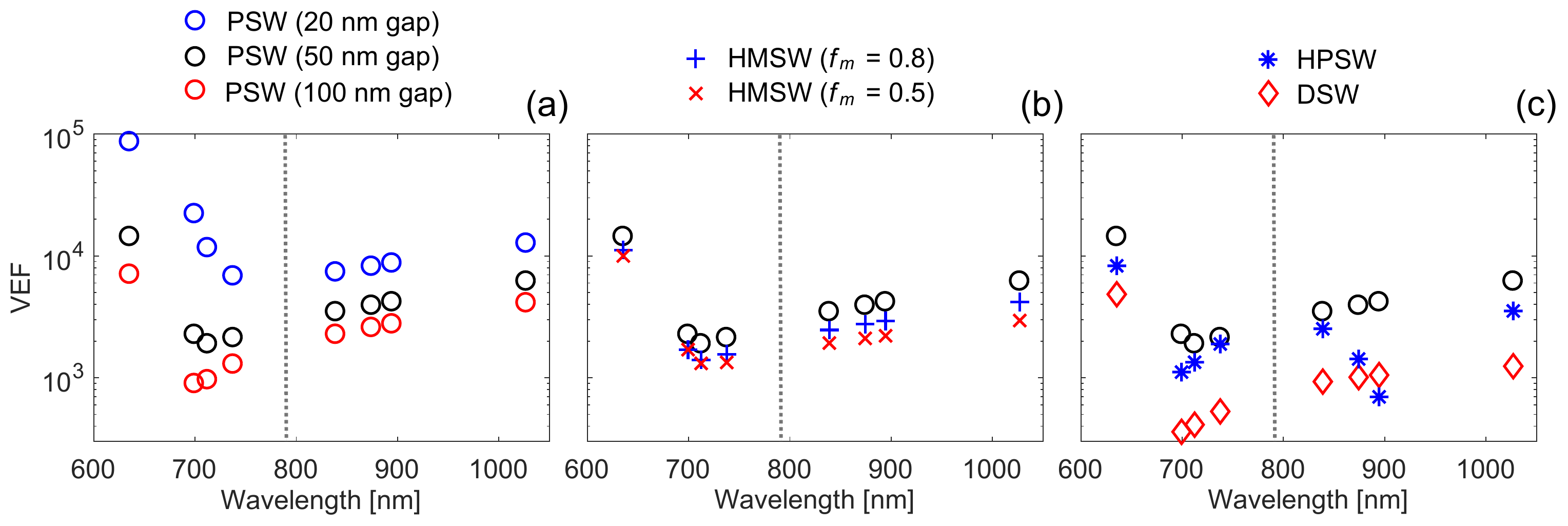}
	\caption{Maximized volume enhancement factor (VEF) with the optimal WG length for the case of in-plane excitation and forward scattered signal collection (see Fig.\,\ref{fig:Accum_EF}), as a function of wavelength for the Ag plasmonic slot WG (PSW) with gap width $g$ = 50 nm, and other WGs obtained by (a) varying the gap width $g$: Ag PSW with gap width $g$ = 100 nm and $g$ = 20 nm, (b) varying the metal filling fraction $f_m$: Ag-$\textnormal{Si}_3\textnormal{N}_4$ HM slot WG (HMSW) with $f_m$ = 0.8 and $f_m$ = 0.5, and (c) changing the WG type: Ag-$\textnormal{Si}_3\textnormal{N}_4$ hybrid plasmonic slot WG (HPSW) and $\textnormal{Si}_3\textnormal{N}_4$ dielectric slot WG (DSW). The results here are based on comparing to a reference Gaussian beam focused using an objective lens with NA = 0.75 that has a waist diameter $D_0 \approx$ 1.28 $\mu\textnormal{m}$ and depth of focus $b \approx$ 17.4 $\mu\textnormal{m}$. Note that changing the Gaussian beam parameters will simply scale the VEF. Each marker indicates a specific Stokes or anti-Stokes wavelength corresponding to one of the Raman modes as described in Section\,\ref{sec:WG_mode_properties}. The vertical dotted line indicates the pump wavelength at $\lambda_0$ = 785 nm.}
	\label{fig:AEF_vs_wavelength}
\end{figure*}

\begin{figure*}
	\centering
	\includegraphics[width=0.95\textwidth]{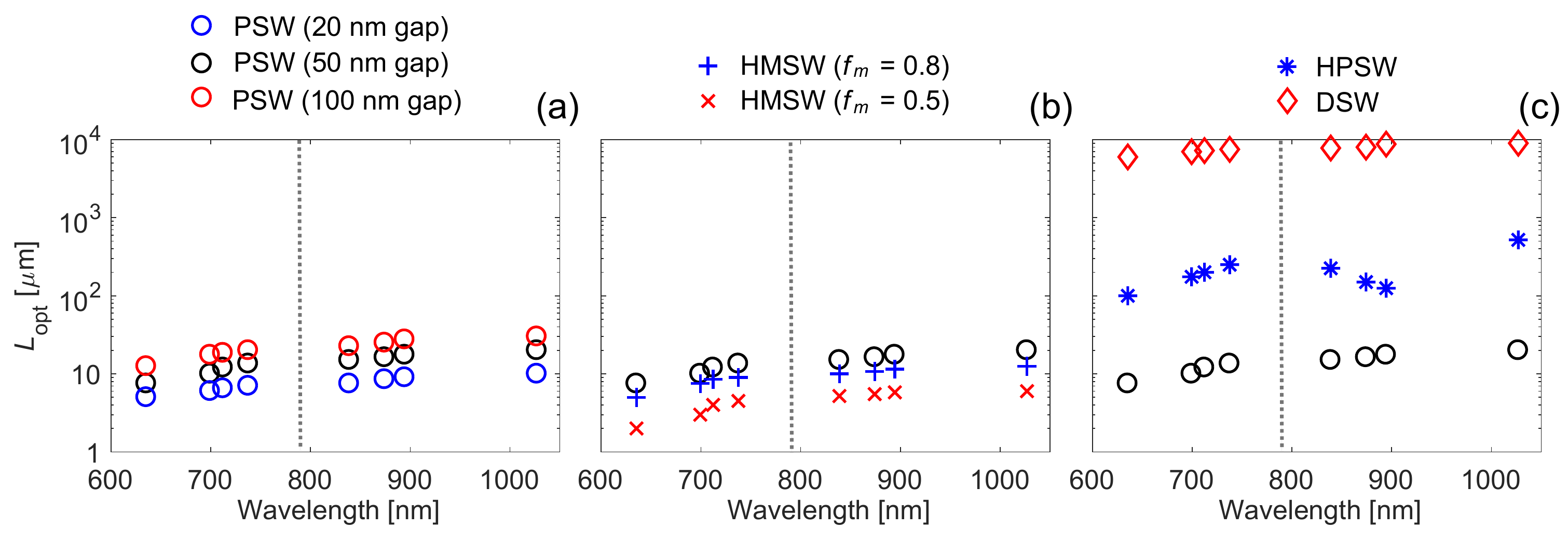}
	\caption{Optimized WG length $L_\textnormal{opt}$ for maximum AEF (see Fig.\,\ref{fig:Accum_EF}) for the case of forward scattered signal collection, as a function of wavelength for the Ag plasmonic slot WG (PSW) with gap width $g$ = 50 nm, and other WGs obtained by (a) varying the gap width $g$: Ag PSW with gap width $g$ = 100 nm and $g$ = 20 nm, (b) varying the metal filling fraction $f_m$: Ag-$\textnormal{Si}_3\textnormal{N}_4$ HM slot WG (HMSW) with $f_m$ = 0.8 and $f_m$ = 0.5, and (c) changing the WG type: Ag-$\textnormal{Si}_3\textnormal{N}_4$ hybrid plasmonic slot WG (HPSW) and $\textnormal{Si}_3\textnormal{N}_4$ dielectric slot WG (DSW). Each marker indicates a specific Stokes or anti-Stokes wavelength corresponding to one of the Raman modes as described in Section\,\ref{sec:WG_mode_properties}. The vertical dotted line indicates the pump wavelength at $\lambda_0$ = 785 nm.}
	\label{fig:L_opt_vs_wavelength}
\end{figure*}

One of the simplest parameters that can be tuned even for the pure metal Ag plasmonic slot WG [Fig.\,\ref{fig:wg_structures}(a)] is the width of the gap region $g$. In Fig.\,\ref{fig:SMEF_vs_wavelength}(a), the maximal SMEF is presented for Ag plasmonic slot WGs with $g$ = 20, 50, and 100 nm. By decreasing $g$ from 50 to 20 nm, the SMEF is increased by over an order of magnitude; at $\lambda_0$ = 839 nm for the Stokes line of Raman mode with $\omega_m$ = 98.7 meV, SMEF increases from 7.58 $\times$ $10^4$ to 7.99 $\times$ $10^5$. When $g$ is increased from 50 to 100 nm, the SMEF conversely decreases to 1.54 $\times$ $10^4$ at $\lambda_0$ = 839 nm. The VEF for the Ag plasmonic slot WG in general follows the same trend as the SMEF, such that by decreasing the gap width $g$, the VEF increases significantly [Fig.\,\ref{fig:AEF_vs_wavelength}(a)]. The magnitude of the change in VEF is highly dependent on the Raman scattered wavelength; when $g$ decreases from 100 to 20 nm, at $\lambda_0$ = 839 nm, VEF increases from 2269 to 7373, but at $\lambda_0$ = 699 nm, the change is by over an order of magnitude from VEF = 895 to 2.22 $\times$ $10^4$. Interestingly, although the VEF for the Ag plasmonic slot WG with $g$ = 20 nm is the highest, the optimal length to achieve the enhancement is the lowest at $L_\textnormal{opt}$ = 7.5 $\mu\textnormal{m}$ for the Raman line at $\lambda_0$ = 839 nm, compared to $L_\textnormal{opt}$ = 22.5 $\mu\textnormal{m}$ for the Ag plasmonic slot WG with $g$ = 100 nm [Fig.\,\ref{fig:L_opt_vs_wavelength}(a)], which is attributed to the higher modal loss as $g$ is decreased. This is an indication that the Raman enhancement contributed by the $\textbf{E}$-field and Purcell enhancement (i.e., SERS-type enhancement) is stronger compared to enhancement by increasing light-matter interaction volume. It was highlighted in Section \ref{subsec:SMEF} that the Raman enhancement of a single molecule is due to both $\textbf{E}$-field enhancement of the excitation light and also Purcell enhancement of the Raman scattered photons. The changes in the magnitudes of these contributions to Raman enhancement as the gap width of the Ag plasmonic slot WG is modified are shown in Figs.\,\ref{fig:Eta_Appendix}(a) and \ref{fig:Fp_vs_wavelength}(a). Clearly, as $g$ is decreased, the maximal $\textbf{E}$-field enhancement factor $|\eta(z_m = 0)|$ increases significantly [Fig.\,\ref{fig:Eta_Appendix}(a)]; also, the Purcell factor $F_x$ increases for all wavelengths (e.g., at $\lambda_0$ = 839 nm, from $F_x$ = 7.6 when $g$ = 100 nm to $F_x$ = 80.1 when $g$ = 20 nm) [Fig.\,\ref{fig:Fp_vs_wavelength}(a)]. The increases in both $|\eta|$ and $F_x$ as the gap width is decreased are attributed to a decrease in energy velocity $\upsilon_\textnormal{E}$, as shown in [Fig.\,\ref{fig:Energy_velocity_vs_wavelength}(a)]. Furthermore, the decrease in gap width leads to a higher EM field concentration inside a smaller cross-sectional area; this contributes to increasing the normalized $\textbf{E}$-field $\textbf{e}(\boldsymbol{\rho})$ within the gap region that is responsible for increasing $\textbf{G}_\textnormal{WG}$ and thus $F_x$ [Eq.\,\eqref{eq:Purcell_factor}], as well as reducing the effective WG mode area $\mathcal{A}_\textnormal{WG}$ that contributes to increasing $|\eta|$ [Eq.\,\eqref{eq:field_enhancement}]. By decreasing the gap width of the Ag plasmonic slot WG, the $\beta$-factor also increases dramatically over all wavelengths, as shown in Fig.\,\ref{fig:beta_vs_wavelength}(a).

From Fig.\,\ref{fig:SMEF_vs_wavelength}(b), it is observed that by changing from a pure metal Ag plasmonic slot WG to stratified HM (Ag-$\textnormal{Si}_3\textnormal{N}_4$) slot WG [Fig.\,\ref{fig:wg_structures}(c)] through reducing the metal filling fraction $f_m$ of the claddings (while keeping the gap width constant), the increase in SMEF is not as pronounced as by reducing the gap width of the pure Ag plasmonic slot WG; at $\lambda_0$ = 893 nm, the SMEF increases to 1.51 $\times$ $10^5$ when $f_m$ is reduced to 0.5. This is attributed to the fact that while one of the contributing factors to Raman enhancement, namely the Purcell factor $F_x$ increases when $f_m$ is reduced [Fig.\,\ref{fig:Fp_vs_wavelength}(b)], the maximal $\textbf{E}$-field enhancement $|\eta(z_m = 0)|$ conversely decreases [Fig.\,\ref{fig:Eta_Appendix}(b)]. The energy velocity $\upsilon_\textnormal{E}$ decreases as $f_m$ is reduced [Fig.\,\ref{fig:Energy_velocity_vs_wavelength}(b)], which indeed contributes to increasing $\textbf{G}_\textnormal{WG}$ and thus $F_x$. However, for the calculation of $|\eta|$, the effective mode area $\mathcal{A}_\textnormal{WG}$ is also important [Eq.\,\eqref{eq:field_enhancement}]. When $f_m$ is reduced, there is an increased EM energy density that resides within the claddings, as shown in Fig.\,\ref{fig:EM_energy_dens}, and hence $\mathcal{A}_\textnormal{WG}$ increases and leads to a decrease of $|\eta|$. An increase of EM field concentration in the lossy HM claddings raises the WG modal loss, and thus the effective propagation length is also reduced as $f_m$ is decreased. Overall, it can be seen in Fig.\,\ref{fig:AEF_vs_wavelength}(b) that the VEF actually decreases as $f_m$ is decreased (at $\lambda_0$ = 839 nm, VEF decreases to 1927 for HM claddings with $f_m$ = 0.5), because the improvement in SMEF is not large enough to compensate for the reduction in light-matter interaction volume due to the reduced optimal WG length $L_\textnormal{opt}$ [Fig.\,\ref{fig:L_opt_vs_wavelength}(b)]. From Fig.\,\ref{fig:beta_vs_wavelength}(b), it is observed that the $\beta$-factor increases as $f_m$ of the HM slot WG claddings is decreased, which follows the same trend as for the SMEF.

Next, we compare the Raman enhancement performance of the Ag plasmonic slot WG with two other types of WG structures, namely the $\textnormal{Si}_3\textnormal{N}_4$ dielectric slot WG and the hybrid plasmonic slot WG, as shown in Figs.\,\ref{fig:wg_structures}(b) and \ref{fig:wg_structures}(d), respectively. Note that the gap width for each WG remains the same at 50 nm. The trends in SMEF and VEF by using different WG types can be observed in Fig.\,\ref{fig:SMEF_vs_wavelength}(c) and \ref{fig:AEF_vs_wavelength}(c), respectively. In general, the Raman enhancement performance of the hybrid plasmonic slot WG is in between that of the Ag plasmonic slot WG and $\textnormal{Si}_3\textnormal{N}_4$ dielectric slot WG. This is attributed to the fact that the contributing figures-of-merit, namely maximal $|\eta|$ [Fig.\,\ref{fig:Eta_Appendix}(c)], $F_x$ [Fig.\,\ref{fig:Fp_vs_wavelength}(c)], and $L_\textnormal{opt}$ [Fig.\,\ref{fig:L_opt_vs_wavelength}(c)] for the hybrid plasmonic slot WG are also in between the respective values of the plasmonic and dielectric slot WGs. As such, the $\beta$-factor for the hybrid plasmonic slot WG also falls between the values of the Ag plasmonic and $\textnormal{Si}_3\textnormal{N}_4$ dielectric slot WGs [Fig.\,\ref{fig:beta_vs_wavelength}(c)]. The use of the hybrid plasmonic slot WG serves the purpose of increasing the effective propagation length (and thus $L_\textnormal{opt}$) that increases light-matter interaction volume compared to the Ag plasmonic WG (by $\sim$15$\times$); however, the SMEF is simultaneously much reduced, e.g., from 7.58 $\times$ $10^4$ to 5.14 $\times$ $10^3$ at $\lambda_0$ = 839 nm.

From Fig.\,\ref{fig:AEF_vs_wavelength}, it can be seen that for each of the different WGs, there exists a minimum VEF at a certain wavelength. This is due to the interplay between the maximum SMEF (SERS-type enhancement) and the $L_\textnormal{opt}$ (enhancement via increase in light-matter interaction volume) achievable with the WG. For wavelengths above the wavelength of minimum VEF, the effective propagation lengths are relatively high due to lower modal losses [Fig.\,\ref{fig:WGs_dispersion}(b)], which contributes to increasing the overall VEF. On the other hand, as the wavelength decreases below that of minimum VEF, although the propagation loss has increased to significantly reduce $L_\textnormal{opt}$ (Fig.\,\ref{fig:L_opt_vs_wavelength}), the SMEF has increased enough to compensate and yield an overall increase in VEF.

By using a purely dielectric WG structure, which we present here using the $\textnormal{Si}_3\textnormal{N}_4$ slot WG as an example [Fig.\,\ref{fig:wg_structures}(b)], the VEF at $\lambda_0$ = 839 nm is 930, which is much lower in comparison to the plasmonic slot WG (VEF = 3480) or the hybrid plasmonic slot WG (VEF = 2520). In our calculations, the propagation loss of the $\textnormal{Si}_3\textnormal{N}_4$ slot WG is assumed to be 10 dB/cm, which is consistent with experimentally determined losses typical for fabricated slot WG structures \cite{BaehrJones2005}. Although the resulting $L_\textnormal{opt}$ for maximal VEF is 7 mm for the $\textnormal{Si}_3\textnormal{N}_4$ slot WG, which is close to 3-orders of magnitude longer than for the Ag plasmonic slot WG, the attainable maximal SMEF is only 46, and thus it exhibits comparatively poor performance in terms of accumulated Raman enhancement on an ensemble of molecules of a bulk sample. We further performed our theoretical calculations on the HC-PCF (Model: NKT HC-800-02 \cite{Mak2013}), and found that by using a 4-cm length of WG, the VEF = 660, which is comparable to that of the $\textnormal{Si}_3\textnormal{N}_4$ dielectric slot WG, although the maximal VEF is 4820 using an optimized HC-PCF length of 75 cm. From Fig.\,\ref{fig:beta_vs_wavelength}(c), it is observed that while plasmonic type slot WGs can achieve $\beta$ $>$ 0.9 (with gap widths of $\leq$50 nm), for dielectric type or hybrid plasmonic slot WGs, the $\beta$-factor is generally $<$0.8. The $\beta$-factors for plasmonic slot WGs shown in this work are comparable to those achievable in photonic crystal WGs \cite{Rao2007_2}. However, the advantage of using plasmonic slot WGs is that high Purcell factors can be maintained over a broad bandwidth of $\sim$400 nm (Fig.\,\ref{fig:Fp_vs_wavelength}), whereas for photonic crystal WGs the bandwidth is typically $<$10 nm\,\,\cite{Rao2007_2}.

% =====================================================================================
\section{Discussion and Conclusions}
\label{sec:Conclusions}

It is observed in Fig.\,\ref{fig:AEF_vs_wavelength} that the best VEF out of all the WGs studied is attained by the Ag plasmonic slot WG with 20 nm width gap, with an impressive value of VEF = 7373 at $\lambda_0$ = 839 nm in a WG length of only 7.5 $\mu\textnormal{m}$. Notably, this value is higher than the VEF of a 75-cm long length of HC-PCF (i.e., 5-orders of magnitude longer). This result emphasizes that in general for plasmonic slot WGs, SERS-type enhancement through the Purcell and $\textbf{E}$-field enhancement contributes more significantly to VEF of a bulk sample compared to enhancement by increased light-matter interaction volume.

By changing from a dielectric WG based on total-internal-reflection to a metal plasmonic WG, the VEF is significantly increased, while the required length of device decreases by close to three-orders of magnitude. It is shown that to further increase the VEF, the strategy is to reduce the gap width; whereas the use of stratified HMs for the claddings or the hybrid plasmonic slot WG does not seem to help further improve Raman enhancement for a bulk sample. However, there may be practical limits to how small the width of the gap can be in terms of filling the gap with the analyte liquid. In previously demonstrated work on dielectric slot WGs for Raman spectroscopy, both $\textnormal{Si}_3\textnormal{N}_4$ and $\textnormal{TiO}_2$ slot WGs with gap widths of 100 to 150 nm have been shown to to effectively enhance the Raman signal beyond that by using strip WGs \cite{Dhakal2015,Rahomaki2013}. The void-free filling of trenches etched in silicon with widths as small as 20 nm (even with depth of $\sim$715 nm and thus aspect ratio of $\sim$32) using spin-on dielectric for shallow trench isolation application in semiconductor processing has also been shown \cite{Trivedi2009}. As such, the filling of nanoscale gaps of the plasmonic slot WGs proposed in this work with analyte for Raman spectroscopy is indeed quite feasible. However, for decreasing gap dimensions, nanofluidics techniques would have to be considered in conjunction with the photonics side of device design \cite{Mukhapadhyay2006,VanDenBerg2005,Duan2013}.

Although plasmonic slot WGs achieve much lower SMEF (on the order of $10^5$ for Ag plasmonic WG with 20 nm gap) compared to metal dimers (also with a 20 nm gap) used in traditional SERS that are on the order of $10^7$\,\,\cite{Wong2017}, here we show that our proposed plasmonic slot WGs can attain Raman enhancement factors close to planar SERS substrates for bulk sample measurements. We borrow the plasmonic dimer structure from Ref. \citenum{Wong2017} for comparison, in which the dimensions of each dimer gap where the hotspot exists are 20 (gap width) $\times$ 35 $\times$ 35 nm, such that the volume of each hotspot is $V_\textnormal{h}$ = 2.45 $\times$ $10^4$ $\textnormal{nm}^3$. By considering that the excitation Gaussian beam waist has a diameter of $\sim$638 nm, only approximately 10 plasmonic dimers on the SERS substrate can be simulataneously excited within the beam cross-section area (Fig.\,\ref{fig:beam_excitation_SERS_dimers}). The achievable VEF is thus calculated by: VEF = SMEF $\times$ 10$V_\textnormal{h}$/$V_\textnormal{0}$, where $V_\textnormal{0}$ is the interrogation volume of the reference case of Gaussian beam excitation of the analyte in free space for which VEF = 1 (i.e., volume of the excitation Gaussian beam waist region). Using SMEF = 4 $\times$ $10^7$ (plasmonic dimer with resonant wavelength of $\sim$785 nm, and for the Stokes line of Raman mode with $\omega_m$ = 98.7 meV at $\lambda_0$ = 839 nm) \cite{Wong2017}, the VEF is $\sim$7000, which is similar to that achievable by using the Ag plasmonic slot WG.

The main advantage of utilizing metal based plasmonic slot WGs for enhanced Raman spectroscopy in bulk sensing would be the potential for integration along with other components such as excitation source and spectrometer on a single planar chip, and also the ultra-low volume of analyte required for sensing ($<$1 fL). Traditional SERS substrates not only has the problem of unstable hotspots and the requirement for functionalization of the metal surface, a bulky setup for excitation and collection of Raman scattered signal is often necessary. Using dielectric WGs such as HC-PCF or even integrated on-chip, the required analyte volume is several orders of magnitude larger, and the maximum Raman enhancement achievable is less than by using a metal plasmonic slot WG. Moreover, the guided mode of the plasmonic slot WG has over 99$\%$ of its power contained in the gap region where light-matter interaction takes place [Fig.\,\ref{fig:EM_energy_dens}(a)], which not only maximizes pump power delivered to the sensing region, but also eliminates any background Raman signal not produced by the analyte molecules. Using a dielectric slot WG, only $\sim$30$\%$ of the mode power resides within the gap region, meaning that the WG material itself generates unwanted background Raman signal. Another attribute of WG based Raman sensing is that the bandwidth over which large Raman enhancement can be maintained is significantly wider compared to SERS on a nanostructured surface. The enhancement factor does not drop below one-order of magnitude lower than the maximum value in a wide bandwidth of $\sim$400 nm (Fig.\,\ref{fig:AEF_vs_wavelength}), whereas for the traditional SERS substrate, this bandwidth is only $\sim$300 nm \cite{Wong2017}. 

In summary, the plasmonic slot WGs proposed in this work applied to enhancing Raman spectroscopy are promising devices for future ultra-compact Raman sensors, with potential for dense integration within on-chip systems containing different technologies including photonics, electronics, and nanofluidics.

\section*{Acknowledgment}

This work was supported by the Natural Sciences and Engineering Research Council of Canada, Queen's University, and Lumerical Solutions Inc. We also thank Simon Axelrod for his discussions.

% =====================================================================================
\appendix

\section{Calculation of Energy Velocity and Waveguide Mode Normalization}
\label{Appendix:energy_velocity}

The energy velocity is given by \cite{Cheng1992}
\begin{equation}
\upsilon_\textnormal{E} = \frac{\langle S_z \rangle}{\langle \mathcal{W} \rangle},
\label{eq:energy_velocity_in_Appendix} 
\end{equation}
where the time-averaged power flow, for a WG mode traveling in the $z$ direction, is
\begin{equation}
\label{eq:power_flow_in_Appendix}
\langle S_z \rangle = \frac{1}{2}\int_S \textnormal{Re}\left[\textbf{E}_{k_\omega}(\boldsymbol{\rho})\times\textbf{H}_{k_\omega}^*(\boldsymbol{\rho})\right]\cdot\hat{z} d\boldsymbol{\rho} 
\end{equation}
and the time-averaged energy per length is in general $\langle \mathcal{W} \rangle = \int_S {\cal w}(\boldsymbol{\rho})d\boldsymbol{\rho}$. The quantity within the integral ${\cal w}(\boldsymbol{\rho}) = {\cal w}_e(\boldsymbol{\rho}) + {\cal w}_h(\boldsymbol{\rho})$ is the energy density that consists of both the electric and magnetic field components as specified by ${\cal w}_e$ and ${\cal w}_h$, respectively. The integration runs over the area of the 2-D simulation region that includes the WG cross-section [Fig.\,\ref{fig:integration_geom}(a)]. For metallic materials that can be described by the Drude-Lorentz medium model, the $\textbf{E}$-field energy density is \cite{Loudon1970,Boardman2006}
\begin{equation}
\label{eq:energy_density_Drude_Lorentz_in_Appendix}
{\cal w}_e = \frac{\varepsilon_0}{4}\left[1 + \frac{\omega_p^2}{\omega^2 + \Gamma^2}\right]|\mathbf{E}_{k_\omega}|^2,
\end{equation}
where $\omega_p$ is the plasma frequency and $\Gamma$ is the damping rate. For a lossless dispersive material, the energy density of the electric field is generalized to \cite{Brillouin1960,Landau1984}
\begin{equation}
\label{eq:energy_density_lossless_in_Appendix}
{\cal w}_e = \frac{\varepsilon_0}{4}\left[\mathbf{E}_{k_\omega}\cdot\frac{\partial(\omega\varepsilon(\omega))}{\partial\omega}\cdot\textbf{E}_{k_\omega}^*\right],
\end{equation}
where $\varepsilon(\omega)$ is in general the frequency-dependent relative permittivity tensor. Based on the Drude-Lorentz medium model for metals, the permittivity is given by
\begin{equation}
\label{eq:Drude_Lorentz}
\varepsilon(\omega) = 1 - \frac{\omega_p^2}{\omega^2 + i\Gamma\omega},
\end{equation}
which is employed to obtain Eq.\,\eqref{eq:energy_density_Drude_Lorentz_in_Appendix}. Since the materials that we are investigating are non-magnetic, the magnetic field energy density is simply $\mathcal{w}_h = 1/4 \mu_0 |\textbf{H}_{k_\omega}|^2$, with the relative permeability $\mu(\omega)$ = 1.

\begin{figure}
	\centering
	\includegraphics[width=\columnwidth]{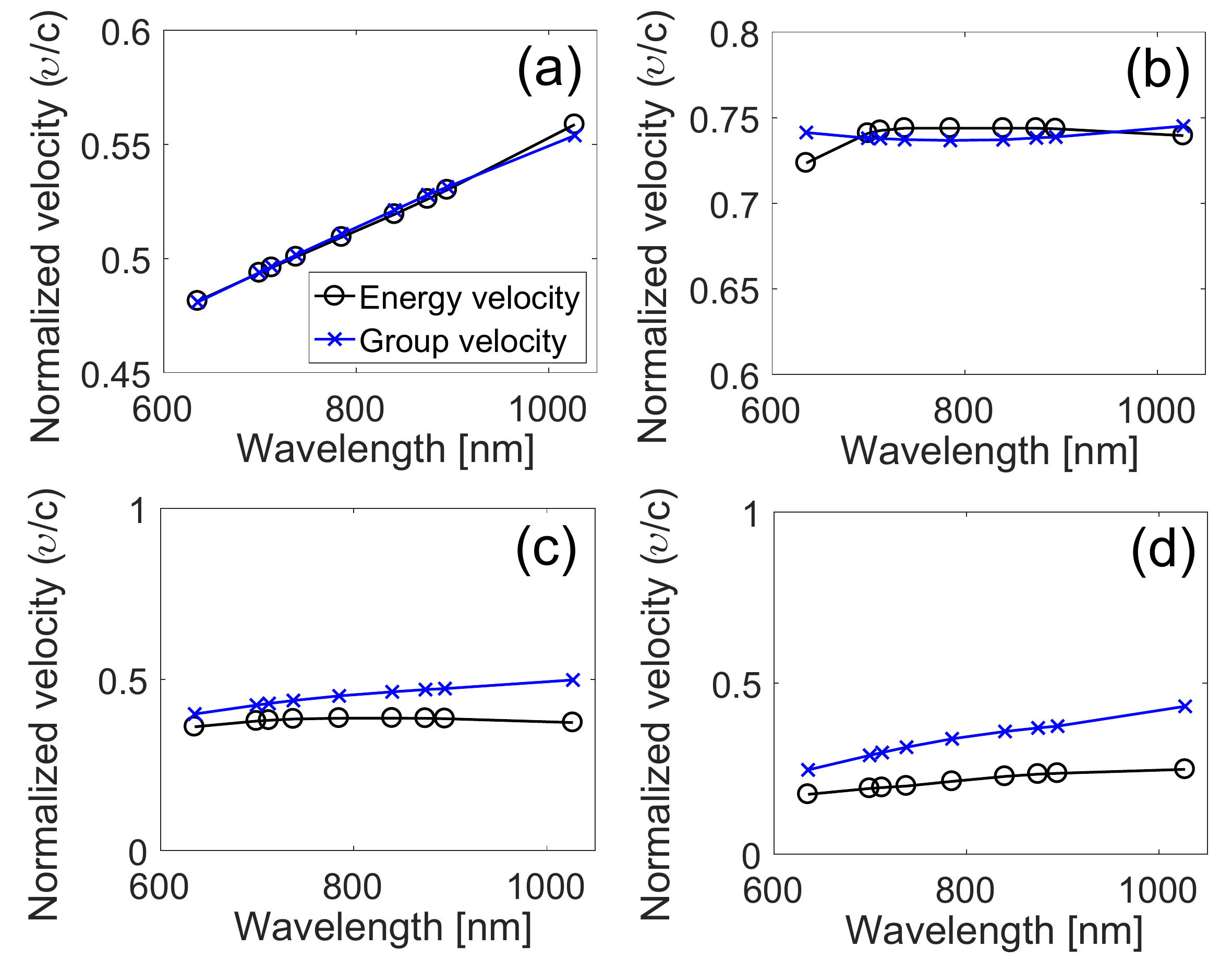}
	\caption{Normalized energy and group velocities ($\upsilon$/c) as a function of wavelength for (a) $\textnormal{Si}_3\textnormal{N}_4$ dielectric slot WG ($\alpha$ = 1.398 $\times$ $10^{-4}$ $\mu\textnormal{m}^{-1}$ at $\lambda$ = 785 nm), (b) hollow-core photonic crystal fiber - Model: NKT Photonics HC-800-02 ($\alpha$ = 2.174 $\times$ $10^{-6}$ $\mu\textnormal{m}^{-1}$ at $\lambda$ = 785 nm), (c) Ag plasmonic slot WG with gap width $g$ = 50 nm ($\alpha$ = 0.0691 $\mu\textnormal{m}^{-1}$ at $\lambda$ = 785 nm), and (d) Ag-$\textnormal{Si}_3\textnormal{N}_4$ HM slot WG with $f_m$ = 0.5 ($\alpha$ = 0.224 $\mu\textnormal{m}^{-1}$ at $\lambda$ = 785 nm).}
	\label{fig:Energy_group_velocities}
\end{figure}

From Fig.\,\ref{fig:Energy_group_velocities}, it can be observed that in general, the energy velocity of a WG mode is different from its group velocity (an approximation based on the assumption of a medium possessing small dispersion \cite{Censor1989}). For a lossless WG, such as the $\textnormal{Si}_3\textnormal{N}_4$ slot WG and the hollow-core photonic crystal fiber, the group velocity is very close to the energy velocity [Fig.\,\ref{fig:Energy_group_velocities}(a) and \ref{fig:Energy_group_velocities}(b)]. However, for a WG exhibiting appreciable loss, such as the Ag plasmonic and HM slot WGs, the energy velocity deviates significantly from the group velocity, and the larger the absorption, the lower the energy velocity is compared to the group velocity [Figs.\,\ref{fig:Energy_group_velocities}(c) and \ref{fig:Energy_group_velocities}(d)].

The normalized modal field $\textbf{e}_{k_\omega}(\boldsymbol{\rho})$ is obtained by first using the commercial mode solver from Lumerical \cite{Lumerical} to capture $\textbf{E}_{k_\omega}(\boldsymbol{\rho})$, and then normalized by the scheme 
\begin{align}
\label{eq:E_field_normalization_in_Appendix}
\mathbf{e}_{k_\omega}(\boldsymbol{\rho}) = 
\varepsilon_0\frac{\mathbf{E}_{k_\omega}(\boldsymbol{\rho})}{\sqrt{\langle \mathcal{W} \rangle}},
\end{align} 
where $\langle \mathcal{W} \rangle$ has been given earlier in this Appendix. This normalization is also discussed in Refs.\citenum{Angelatos2016} and \citenum{Yao2010}, but here it is generalized for lossy and dispersive media.

% =========================================================================================

\section{Comparison of Different Waveguide Green Function Formulations}
\label{Appendix:WG_Green_function}

The Green function $\textbf{G}_\textnormal{WG}$ as shown in Eq.\,\eqref{eq:Green_function} is applicable to WGs that are inherently lossy. Also note that, as specified in Ref.\,\citenum{Angelatos2015}, this form of the Green function includes an additional factor of $\omega^2/c^2 = k_0^2$ relative to other common sources \cite{Novotny2012}, since we use the following source equation:
\begin{equation}
\label{eq:Helmholtz_eq}
\nabla \times \nabla \times \mathbf{G}(\textbf{r},\mathbf{r}^\prime;\omega) - \frac{\omega^2}{c^2}\varepsilon(\mathbf{r},\omega)\mathbf{G}(\textbf{r},\mathbf{r}^\prime;\omega) = \frac{\omega^2}{c^2}\mathbf{I}\delta(\mathbf{r} - \mathbf{r}^\prime). 
\end{equation}
An alternative form of the WG Green function is \cite{Cano2013} 
\begin{equation}
\mathbf{G}_\textnormal{WG,1}(\textbf{r},\mathbf{r}^\prime;\omega) = -\frac{\mathbf{E}_{k_\omega}(\boldsymbol{\rho})\mathbf{E}_{k_\omega}^*(\boldsymbol{\rho}^\prime)e^{i\tilde{k}(z-z^\prime)}}{2i\omega\mu_0 {\cal S}},
\label{eq:Green_function_2_in_Appendix} 
\end{equation}
where ${\cal S} = \int_S \textnormal{Re}\left[\textbf{E}_{k_\omega}(\boldsymbol{\rho})\times\textbf{H}_{k_\omega}^*(\boldsymbol{\rho})\right]\cdot\hat{z} d\boldsymbol{\rho}$, and the integral is over the 2-D simulation area that contains the WG cross-section [Fig.\,\ref{fig:integration_geom}(a)].

The equivalence between Eq.\,\eqref{eq:Green_function} and Eq.\,\eqref{eq:Green_function_2_in_Appendix} is presented here. The Green function of the forward propagating mode ($z > z'$) is 
\begin{equation}
\mathbf{G}_\textnormal{WG,2}(\textbf{r},\mathbf{r}^\prime;\omega) = \frac{i\omega}{2\upsilon_\textnormal{E}}\mathbf{e}_{k_\omega}(\boldsymbol{\rho})\mathbf{e}_{k_\omega}^*(\boldsymbol{\rho}^\prime)e^{i\tilde{k}(z-z^\prime)}.
\label{eq:Green_function_forward_no_loss_in_Appendix} 
\end{equation}
Without WG loss, the time-averaged energy per length can be calculated by utilizing Eq.\,\eqref{eq:energy_density_lossless_in_Appendix}, and it becomes
\begin{align}
\langle \mathcal{W} \rangle =   &\frac{1}{4}\int_{S}\bigg[\varepsilon_0\mathbf{E}_{k_\omega}(\boldsymbol{\rho})\cdot\frac{\partial(\omega\varepsilon(\boldsymbol{\rho},\omega))}{\partial\omega}\cdot\textbf{E}_{k_\omega}^*(\boldsymbol{\rho}) \nonumber \\
&+ \mu_0\textbf{H}_{k_\omega}(\boldsymbol{\rho})\cdot\textbf{H}_{k_\omega}^*(\boldsymbol{\rho})\bigg]d\boldsymbol{\rho}.
\label{eq:energy_per_length_lossless_in_Appendix} 
\end{align}
Also, the time-averaged power flow is simply given by $\langle S_z \rangle = \frac{1}{2}{\cal S}$. The Green function expression in Eq.\,\eqref{eq:Green_function_forward_no_loss_in_Appendix} can be expanded by using Eqs.\,\eqref{eq:energy_velocity_in_Appendix} and \eqref{eq:E_field_normalization_in_Appendix}, which becomes
\begin{equation}
\mathbf{G}_\textnormal{WG,2}(\textbf{r},\mathbf{r}^\prime;\omega) = \frac{i\omega\varepsilon_0\mathbf{E}_{k_\omega}(\boldsymbol{\rho})\mathbf{E}_{k_\omega}^*(\boldsymbol{\rho}^\prime)e^{i\tilde{k}(z-z^\prime)}}{2 {\cal S}}.
\label{eq:Green_function_expanded_in_Appendix} 
\end{equation}
By using the identity $c = 1/\sqrt{\varepsilon_0\mu_0}$, the expression for the WG Green function would be
\begin{equation}
\mathbf{G}_\textnormal{WG,2}(\textbf{r},\mathbf{r}^\prime;\omega) = -\frac{k_0^2\mathbf{E}_{k_\omega}(\boldsymbol{\rho})\mathbf{E}_{k_\omega}^*(\boldsymbol{\rho}^\prime)e^{i\tilde{k}(z-z^\prime)}}{2i\omega\mu_0 {\cal S}},
\label{eq:Green_function_final_in_Appendix}
\end{equation}
which is equivalent to Eq.\,\eqref{eq:Green_function_2_in_Appendix} with an additional factor $k_0^2$, as expected.

% =======================================================================

\section{Derivation of the Field Enhancement Factor}
\label{Appendix:field_enhancement}

In the calculation of the $\textbf{E}$-field enhancement $|\eta(z_m)|$ presented here, which can be used to determine the effective excitation amplitude seen by the molecule in the presence of the WG, it is assumed that there is no coupling loss from the Gaussian beam to the WG mode, such that the power of the Gaussian beam $P_0$ is completely transferred to become the WG mode power at the input facet $P_\textnormal{WG,0}$. The details of coupling loss are treated separately as the result would just be an efficiency value that serves to decrease $P_\textnormal{WG,0}$. Due to the lossy nature of the WGs investigated, the power within the WG would also decrease as a function of propagation distance away from the input facet, such that the WG power becomes $P_\textnormal{WG}(z_m) = P_\textnormal{WG,0}e^{-\alpha_P z_m}$, where $\alpha_p = 2\kappa_p$ is the WG modal loss at the pump wavelength, and thus $|\eta|$ would also be dependent on the molecule location along the WG $z_m$.

From Eq.\,\eqref{eq:energy_velocity_in_Appendix}, it can be deduced that the WG mode energy per unit length is given by $\langle \mathcal{W}_\textnormal{WG} \rangle = \langle S_z^\textnormal{WG} \rangle / \upsilon_\textnormal{E} = P_\textnormal{WG}/\upsilon_\textnormal{E}$. Similarly, for the Gaussian beam that is propagating at a speed $c/n_\textnormal{B}$ within the background medium, the energy per unit length is $\langle \mathcal{W}_0 \rangle = P_0 n_\textnormal{B}/c$. From Eq.\,\eqref{eq:energy_per_length_lossless_in_Appendix}, it is seen that the approximation $\langle \mathcal{W}_\textnormal{WG} \rangle = 1/4\varepsilon_0\int_{S}|\textbf{E}_\textnormal{WG}(\boldsymbol{\rho})|^2d\boldsymbol{\rho}$ can be made, where the ``WG" subscript in $\textbf{E}_\textnormal{WG}$ implicitly specifies that it is the modal $\textbf{E}$-fields. Similarly, it is estimated that $\langle \mathcal{W}_0 \rangle = 1/4\varepsilon_0\int_{S}|\textbf{E}_0(\boldsymbol{\rho})|^2d\boldsymbol{\rho}$. By the Mean Value Theorem for integrals \cite{Stewart2003} [also shown in Eq.\,\eqref{eq:Mean_value_theorem_in_Appendix}], $\langle \mathcal{W}_\textnormal{WG} \rangle = 1/4\varepsilon_0|\textbf{E}_\textnormal{WG,avg}|^2 \mathcal{A}_\textnormal{WG}$ and $\langle \mathcal{W}_0 \rangle = 1/4\varepsilon_0|\textbf{E}_\textnormal{0,avg}|^2 \mathcal{A}_0$, where $|\textbf{E}_\textnormal{WG,avg}|$ and $|\textbf{E}_\textnormal{0,avg}|$ are the average $\textbf{E}$-field amplitudes of the WG mode and the Gaussian beam waist, respectively. Also, $\mathcal{A}_\textnormal{WG}$ and $\mathcal{A}_0$ are the effective WG mode area and the area of the Gaussian beam waist, respectively. The field enhancement can thus be formulated as follows:
\begin{align}
|\eta(z_m)|^2 &= \frac{|\textbf{E}_\textnormal{WG,avg}|^2}{|\textbf{E}_\textnormal{0,avg}|^2} = \frac{P_\textnormal{WG}(z_m)}{P_0}\frac{c}{n_\textnormal{B}\upsilon_\textnormal{E}}\frac{\mathcal{A}_0}{\mathcal{A}_\textnormal{WG}} \nonumber \\
&= \frac{P_\textnormal{WG,0}}{P_0}\frac{c}{n_\textnormal{B}\upsilon_\textnormal{E}}\frac{\mathcal{A}_0}{\mathcal{A}_\textnormal{WG}}e^{-\alpha_P z_m}.
\label{eq:field_enhancement_formulation_in_Appendix} 
\end{align}
Since it is assumed that there is no coupling loss from the Gaussian beam to the WG mode, $P_\textnormal{WG,0} = P_0$, and thus we arrive at
\begin{equation}
\label{eq:field_enhancement_in_Appendix}
|\eta(z_m)|^2 = \frac{c}{n_\textnormal{B}\upsilon_\textnormal{E}}\frac{\mathcal{A}_0}{\mathcal{A}_\textnormal{WG}}e^{-\alpha_p z_m}.
\end{equation}
Due to the utilization of the average $\textbf{E}$-field amplitudes of the WG mode and the Gaussian beam waist in Eq.\,\eqref{eq:field_enhancement_formulation_in_Appendix}, $|\eta|$ is only dependent on molecule location along the WG length $z_m$, but not its location within the WG cross-section $\boldsymbol{\rho}_m$.

Lastly, the effective WG mode area is given by
\begin{equation}
\mathcal{A}_\textnormal{WG} = \frac{1}{\textnormal{max}\{{\cal w}(\boldsymbol{\rho})\}}\int_{\mathcal{A}_\textnormal{WG}}{\cal w}(\boldsymbol{\rho})d\boldsymbol{\rho},
\label{eq:effective_mode_area_in_Appendix} 
\end{equation}
where $\textnormal{max}\{{\cal w}(\boldsymbol{\rho})\}$ is the maximum WG mode energy density ${\cal w}(\boldsymbol{\rho})$ at a certain cross-section location $\boldsymbol{\rho}$ within the WG mode area.

% =========================================================================

\section{Accounting for Propagation Effects and Integration Geometries for Different Raman Signal Collection Configurations}
\label{Appendix:Calc_G}

\begin{figure*}
        \centering
        \includegraphics[width=0.8\textwidth]{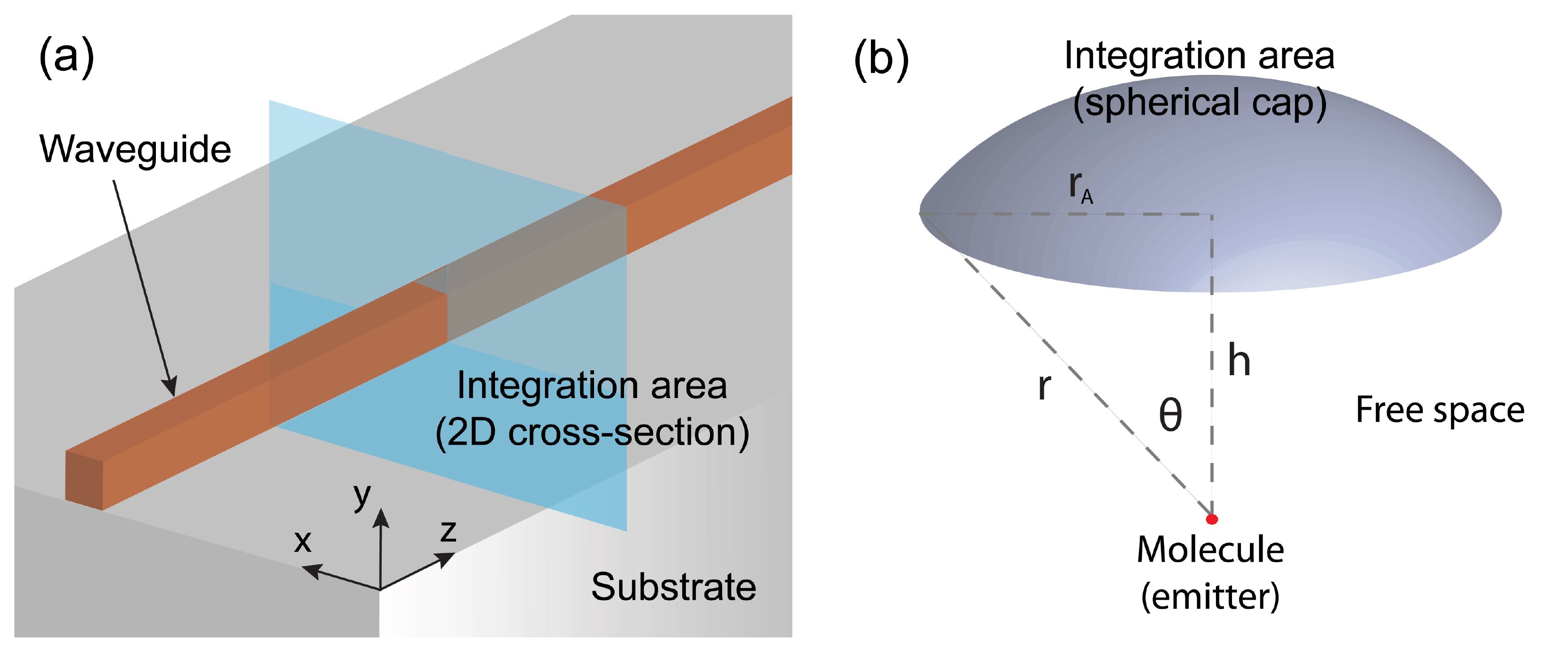}
        \caption{Representation of the integration geometry for obtaining Raman scattered power when the molecule of interest is (a) in the vicinity of a WG modeled using Lumerical MODE \cite{Lumerical}, and (b) when in free space treated analytically, respectively.}
        \label{fig:integration_geom}
\end{figure*}

For the cases of Raman scattered light in the forward or backward direction coupled to the WG mode [Fig.\,\ref{fig:exp_config}(a)], the calculation of $|\mathbf{G}(\mathbf{r}_\textnormal{D},\mathbf{r}_m;\omega) \cdot \mathbf{n}|^2$ requires the use of the WG Green function $\mathbf{G}_\textnormal{WG}(\mathbf{r},\mathbf{r}';\omega)$ as shown in Eq.\,\eqref{eq:Green_function}. In these cases, the molecule location $\textbf{r}_m$ can be separated into the location within the WG cross-section, $\boldsymbol{\rho}_m$, and the location along the length of the WG, $z_m$. Similarly, the detection location $\textbf{r}_\textnormal{D}$ can be separated into $\boldsymbol{\rho}_\textnormal{D}$ and $z_\textnormal{D}$. For Raman scattering in the forward direction to the end facet of the WG, the relevant Green function is $\textbf{G}_\textnormal{WG}(\boldsymbol{\rho}_\textnormal{D},z=L;\boldsymbol{\rho}_m,z=z_m;\omega)$, where $L$ is the length of the WG; and for Raman scattering in the backward direction to the input facet of the WG, the Green function would be $\textbf{G}_\textnormal{WG}(\boldsymbol{\rho}_\textnormal{D},z=0;\boldsymbol{\rho}_m,z=z_m;\omega)$. The integration geometry to calculate $P^\textnormal{SM}$ in Eq.\,\eqref{eq:SERS_power} for both the cases of forward and backward Raman scattering is thus the 2-D cross-section area that encompasses the WG mode, as shown in Fig.\,\ref{fig:integration_geom}(a).

For the case of Raman scattering in the out-of-plane direction [Fig.\,\ref{fig:exp_config}(a)], the calculation of $|\mathbf{G}(\mathbf{r}_\textnormal{D},\mathbf{r}_m;\omega) \cdot \mathbf{n}|^2$ requires the use of the free space Green function $\mathbf{G}_\textnormal{B}(\mathbf{r},\mathbf{r}';\omega)$, which is obtained analytically and only depends on the position $\textbf{r}$ with respect to the molecule location \cite{Ru2007}. The Raman scattered power $P^\textnormal{SM}$ is thus calculated with the integration area $\mathcal{A}_\textnormal{D}$ corresponding to that of the spherical cap centered at the molecule location and with angle $\theta$ set to correspond to the numerical aperture of the WG mode: NA = $n_\textnormal{B}\sin\theta$. A schematic of the integration geometry is shown in Fig.\,\ref{fig:integration_geom}(b), and the spherical cap surface area is given by $\mathcal{A}_\textnormal{sph.cap.} = 2\pi r^2(1 - \textnormal{cos}\theta)$, where $r$ can be arbitrarily set, because an increase in $r$ increases $\mathcal{A}_\textnormal{sph.cap.}$ but simultaneously decreases the Raman scattered power per area as $\mathbf{G}_\textnormal{B}$ decays as a function of $r$. Similarly, $\mathbf{G}_\textnormal{B}$ and the spherical cap integration geometry as shown in Fig.\,\ref{fig:integration_geom}(b) are employed in the calculation of $|\mathbf{G}(\mathbf{r}_\textnormal{D},\mathbf{r}_m;\omega) \cdot \mathbf{n}|^2$ and thus the Raman scattered power for a molecule in free space, $P_0^\textnormal{SM}$.

% ============================================================================

\section{Derivation of Spatially Averaged Raman Enhancement Factor}
\label{Appendix:accum_enhancement}

Commonly in Raman spectroscopy, an ensemble of molecules is excited and the accumulated Raman scattered light from them is measured. A WG serves to increase significantly the number of molecules that interact with the pump light as it propagates along the WG via the mode that it is coupled to. The molecules in the vicinity of the WG are at different locations, in terms of both the cross-section position $\boldsymbol{\rho}_m$ and the position along the WG length $z_m$, which means that the Raman scattered power from each molecule $P^\textnormal{SM}$ is different due to the position-dependent $\textbf{E}$-field amplitude and WG modal loss, as detailed in Section \ref{subsec:SMEF}.

The spatially averaged Raman enhancement factor (AEF) is given in Eq.\,\eqref{eq:AEF} and it is reiterated here:
\begin{equation}
\label{eq:AEF_in_Appendix}
\textnormal{AEF}(L) = \frac{\int_{z_m=0}^{L}\int_{\mathcal{A}_m}P^\textnormal{SM}(\mathbf{r}_m)d\boldsymbol{\rho}_mdz_m}{P^\textnormal{SM}_0 \mathcal{A}_m L},
\end{equation}
which is dependent on both the excitation and Raman signal collection configurations, because the calculation of $P^\textnormal{SM}$ as described in Subsection\,\ref{subsec:SMEF} is also dependent on the way in which pump light is coupled into the WG and Raman scattered light is collected. The Raman scattered power of a single molecule detected in the forward, backward, and out-of-plane directions are denoted as $P^\textnormal{SM}_\textnormal{F}$, $P^\textnormal{SM}_\textnormal{B}$, and $P^\textnormal{SM}_\textnormal{O}$, respectively. Note that in this formulation, we ignore the effects of molecule interactions as the influence of dipole-dipole coupling is in general negligible for liquid samples \cite{Chew1983}. As well, due to the low pump intensities typically utilized in Raman spectroscopy \cite{Mak2013}, which we also adopt in our calculations, the effect of stimulated Raman scattering would not be present \cite{Gazengel1979,Gorelik2015}, and thus it is not considered in our calculations. 

In order to determine the AEF as a function of WG length more easily, it is necessary to decompose the integrand in Eq.\,\eqref{eq:AEF_in_Appendix} into its components; one can do this by taking a closer look at the formulation of $P^\textnormal{SM}$ in Eq.\,\eqref{eq:SERS_power}, and in particular the different components of $S(\textbf{r}_\textnormal{D},\textbf{r}_m,\omega)$ as shown in Eq.\,\eqref{eq:Scat_spect}. First, the prefactor $B$ in $S(\textbf{r}_\textnormal{D},\textbf{r}_m,\omega)$ is only dependent on $z_m$ but not on the cross-sectional position $\boldsymbol{\rho}_m$. Although the second term of $S(\textbf{r}_\textnormal{D},\textbf{r}_m,\omega)$, namely $S_0(\mathbf{r}_m,\omega)$, is a function of the molecule location $\textbf{r}_m$ through $\textnormal{Im}\{\mathbf{G}_{\textnormal{WG},nn}(\mathbf{r}_m,\mathbf{r}_m;\omega)\}$, actually $J_\textnormal{ph}$ is negligible in comparison to $\gamma_m \bar{n}^\textnormal{th}$ for typical excitation intensities on the order of 10s of $\textnormal{mW/cm}^2$, and thus $\textnormal{Im}\{\mathbf{G}_{\textnormal{WG},nn}(\mathbf{r}_m,\mathbf{r}_m;\omega)\}$ in general does not have much effect on the value of $S_0(\mathbf{r}_m,\omega)$. This means that $S_0(\omega)$ is close to a constant value as a function of spatial position and can be factored out of the integrals in Eq.\,\eqref{eq:AEF_in_Appendix} as a valid approximation. For simplicity, it is evaluated only at the position with $z_m = 0$ and $\boldsymbol{\rho}_m$ at the center of the gap region in a slot type WG. 

Finally, the remaining term in Eq.\,\eqref{eq:Scat_spect} is $|\mathbf{G}(\mathbf{r}_\textnormal{D},\mathbf{r}_m;\omega) \cdot \mathbf{n}|^2$, which is dependent on both the molecule location $\textbf{r}_m$ and the detection location $\textbf{r}_\textnormal{D}$. The molecule location $\textbf{r}_m$ is determined by both the position along the WG $z_m$ and the position within the cross-section $\boldsymbol{\rho}_m$. In the case of forward scattering along the WG, the detection location $\textbf{r}_\textnormal{D}$ is defined with $z_m = L$ (at the end facet) and $\boldsymbol{\rho}_\textnormal{D}$ being the range of all cross-section locations; in the backward scattering case, $z_m = 0$ (at the input facet). For out-of-plane scattering, $\textbf{r}_\textnormal{D}$ represents the range of locations defined by the spherical cap area as shown in Fig.\,\ref{fig:integration_geom}(b). More details of the detection area calculation for out-of-plane scattering is given in Appendex\,\ref{Appendix:Calc_G}. The full spatial integration of this term, namely $|\mathbf{G}(\mathbf{r}_\textnormal{D},\mathbf{r}_m;\omega)\cdot\mathbf{n}|^2$ for the forward scattering case, running over the set of possible molecule locations along the WG and within the cross-sectional area, and also the detection area can be expressed as 
\begin{align}
& \textnormal{Spat. Int.} = \nonumber \\
& \frac{\omega^2}{4\upsilon_\textnormal{E}^2}\int_{z_m=0}^L e^{-\alpha_R(L - z_m)}\int_{\mathcal{A}_m}|\textbf{e}_{k_\omega,n}(\boldsymbol{\rho}_m)|^2d\boldsymbol{\rho}_m \nonumber \\
& \int_{\mathcal{A}_\textnormal{D}}|\textbf{e}_{k_\omega,n}(\mathbf{r}_\textnormal{D})|^2d\mathbf{r}_\textnormal{D},
\label{eq:Spat_int_propagator_in_Appendix} 
\end{align}
where $\alpha_R = 2\kappa_R$ is the WG modal loss at the Raman scattered wavelength, and $\textbf{e}_{k_\omega,n}$ is the component of the normalized WG modal complex $\textbf{E}$-field in the direction $\textbf{n}$. As such, it can be seen that the integral running over $\mathcal{A}_m$ in Eq.\,\eqref{eq:AEF_in_Appendix} only needs to be applied to the term $|\textbf{e}_{k_\omega,n}(\boldsymbol{\rho}_m)|^2$, and the integral along the WG from $z = z_m$ to the end facet at $z = L$ is isolated to the term $e^{-\alpha_R(L - z_m)}$. This means that the calculation for the AEF as given in Eq.\,\eqref{eq:AEF_in_Appendix} can be much simplified and more intuitive.

By applying the Mean Value Theorem for integrals \cite{Stewart2003} to the integration that runs over $\mathcal{A}_m$ in Eq.\,\eqref{eq:Spat_int_propagator_in_Appendix}, it can be seen that in fact the average normalized modal $\textbf{E}$-field $\textbf{e}_{k_\omega,\textnormal{avg}}$ can be utilized, such that 
\begin{equation}
\int_{\mathcal{A}_m}|\textbf{e}_{k_\omega,n}(\boldsymbol{\rho}_m)|^2d\boldsymbol{\rho}_m = |\textbf{e}_{k_\omega,\textnormal{avg},n}|^2\mathcal{A}_m,
\label{eq:Mean_value_theorem_in_Appendix} 
\end{equation}
where $\textbf{e}_{k_\omega,\textnormal{avg},n}$ is the component of the average normalized $\textbf{E}$-field in the $\textbf{n}$-direction, and the average normalized $\textbf{E}$-field is given by $\textbf{e}_{k_\omega,\textnormal{avg}} = 1/\mathcal{A}_m\int_{\mathcal{A}_m}\textbf{e}_{k_\omega}(\boldsymbol{\rho}_m)d\boldsymbol{\rho}_m$. In this way, the AEF in the forward direction in Eq.\,\eqref{eq:AEF_in_Appendix} can now be expressed as 
\begin{align}
\label{eq:Ram_em_power_forward_mod_in_Appendix}
\textnormal{AEF}_\textnormal{F}(L) &= \frac{\int_{z_m=0}^{L}P_\textnormal{F,avg}^\textnormal{SM}(z_m)\mathcal{A}_mdz_m}{P^\textnormal{SM}_0 \mathcal{A}_m L} \nonumber \\  
&= \frac{\int_{z_m=0}^{L}P_\textnormal{F,avg}^\textnormal{SM}(z_m)dz_m}{P^\textnormal{SM}_0 L},
\end{align}
where $P_\textnormal{F,avg}^\textnormal{SM}(z_m)$ is the average Raman scattered power per single molecule in the forward direction that reaches the end facet; this is also calculated by Eq.\,\eqref{eq:SERS_power}, but now employing the average normalized $\textbf{E}$-field $\textbf{e}_{k_\omega,\textnormal{avg}}$ (except for the integral in Eq.\,\eqref{eq:Spat_int_propagator_in_Appendix} that runs over $\mathcal{A}_\textnormal{D}$), and it is assumed that the molecule location within the WG cross-section $\boldsymbol{\rho}_m$ is at the center of the gap region (for a slot type WG).

Further simplication is obtained by factoring out the $z$-dependent terms in $P_\textnormal{F,avg}^\textnormal{SM}(z_m)$ [i.e., within Eqs.\,\eqref{eq:field_enhancement_in_Appendix} and \eqref{eq:Spat_int_propagator_in_Appendix}], so that
\begin{equation}
P_\textnormal{F,avg}^\textnormal{SM}(z_m) = P_\textnormal{F,avg}^\textnormal{SM}(z_m=0,L\rightarrow 0) e^{-\alpha_\textnormal{P}z_m} e^{-\alpha_\textnormal{R}(L - z_m)},
\label{eq:Ram_em_power_forward_2_in_Appendix} 
\end{equation}
in which $\alpha_\textnormal{P} = 2\kappa_\textnormal{P}$ and $\alpha_\textnormal{R} = 2\kappa_\textnormal{R}$ are the modal losses at the pump and Raman scattered wavelengths, respectively. The term $P_\textnormal{F,avg}^\textnormal{SM}(z_m=0,L\rightarrow 0)$ represents the Raman scattered power in the limiting case when the molecule is at the input facet and the length of the WG approaches zero, which is denoted $P_\textnormal{F,avg,0}^\textnormal{SM}$. Equation\,\eqref{eq:Ram_em_power_forward_2_in_Appendix} is valid when the pump $\textbf{E}$-field amplitude $\textbf{E}_0$ is sufficiently low such that $J_\textnormal{ph}$ is too small to significantly influence the value of $S_0$ [Eqs.\,\eqref{eq:S0_st} and \eqref{eq:S0_as}]. Similarly, the average Raman signal power from a molecule detected in the backward scattered direction is given as $P_\textnormal{B,avg}^\textnormal{SM}(z_m) = P_\textnormal{B,avg,0}^\textnormal{SM} e^{-\alpha_\textnormal{P}z_m} e^{-\alpha_\textnormal{R}z_m}$, where $P_\textnormal{B,avg,0}^\textnormal{SM} = P_\textnormal{B,avg}^\textnormal{SM}(z_m=0,L\rightarrow 0)$. As such, the integral in Eq.\,\eqref{eq:Ram_em_power_forward_mod_in_Appendix} can be evaluated analytically, and thus the AEF for the cases of detection in the forward and backward directions along the WG are, respectively,
\begin{align}
\label{eq:forward_scat_tot_power_in_Appendix}
& \textnormal{AEF}_{\textnormal{F}}(L) = \\
& \frac{P_\textnormal{F,avg,0}^\textnormal{SM}}{P_0^\textnormal{SM}} \frac{e^{-\alpha_RL}}{L}\frac{1}{\alpha_R -\alpha_P}\left[e^{(\alpha_R - \alpha_P)L} - 1\right], \nonumber \\
& \nonumber \\
\label{eq:backward_scat_tot_power_in_Appendix}
& \textnormal{AEF}_{\textnormal{B}}(L) = \\
& \frac{P_\textnormal{B,avg,0}^\textnormal{SM}}{P_0^\textnormal{SM}} \frac{1}{L(\alpha_R + \alpha_P)}\left[1 - e^{-(\alpha_R + \alpha_P)L}\right]. \nonumber
\end{align} 
For the case of out-of-plane scattering, the last term in Eq.\,\eqref{eq:Scat_spect} is not dependent on the WG loss, as the Raman scattered light is propagated in free space. As such, the AEF detected in the out-of-plane direction with respect to the WG becomes
\begin{equation}
\label{eq:out_scat_tot_power_in_Appendix}
\textnormal{AEF}_{\textnormal{O}}(L) = \frac{P_\textnormal{O,avg,0}^\textnormal{SM}}{P_0^\textnormal{SM}}\frac{1 - e^{-\alpha_P L}}{\alpha_P L}, 
\end{equation}
where $P_{\textnormal{O,avg,0}}^\textnormal{SM} = P_\textnormal{O,avg}^\textnormal{SM}(z_m=0,L\rightarrow 0)$. The average single-molecule enhancement factor can then be defined as $\textnormal{SMEF}_\textnormal{avg} = P_\textnormal{avg}^{\textnormal{SM}}/P_0^{\textnormal{SM}}$ [similar to in Eq.\,\eqref{eq:SMEF}], where $P_\textnormal{avg}^{\textnormal{SM}}$ can either be $P_\textnormal{F,avg}^{\textnormal{SM}}$, $P_\textnormal{B,avg}^{\textnormal{SM}}$ or $P_\textnormal{O,avg}^{\textnormal{SM}}$ depending on the collection configuration of Raman scattering [Fig.\,\ref{fig:exp_config}(a)]. 

From Eqs.\,\eqref{eq:forward_scat_tot_power_in_Appendix}, \eqref{eq:backward_scat_tot_power_in_Appendix}, and \eqref{eq:out_scat_tot_power_in_Appendix}, the spatially averaged Raman enhancement factor (AEF) in the forward, backward, and out-of-plane scattered directions become, respectively,
\begin{align}
\textnormal{AEF}_\textnormal{F}(L) &= \textnormal{SMEF}_\textnormal{F,avg,0}\,e^{-\alpha_RL}\frac{1}{L}\left[\frac{e^{(\alpha_R - \alpha_P)L} - 1}{\alpha_R - \alpha_P}\right], 
\label{eq:EF_forward_2_in_Appendix}
\\  
\textnormal{AEF}_\textnormal{B}(L) &= \textnormal{SMEF}_\textnormal{B,avg,0}\frac{1}{L}\left[\frac{1 - e^{-(\alpha_R + \alpha_P)L}}{\alpha_R + \alpha_P}\right],
\label{eq:EF_backward_2_in_Appendix}
\\  
\textnormal{AEF}_\textnormal{O}(L) &= \textnormal{SMEF}_\textnormal{O,avg,0}\frac{1}{L}\left[\frac{1 - e^{-\alpha_P L}}{\alpha_P}\right], 
\label{eq:EF_out_2_in_Appendix}
\end{align} 
where $\textnormal{SMEF}_\textnormal{F,avg,0} = \textnormal{SMEF}_\textnormal{F,avg}(z_m=0,L\rightarrow 0)$, $\textnormal{SMEF}_\textnormal{B,avg,0} = \textnormal{SMEF}_\textnormal{B,avg}(z_m=0,L\rightarrow 0)$, and $\textnormal{SMEF}_\textnormal{O,avg,0} = \textnormal{SMEF}_\textnormal{O,avg}(z_m=0,L\rightarrow 0)$. Note that in each of Eqs.\,\eqref{eq:EF_forward_2_in_Appendix}, \eqref{eq:EF_backward_2_in_Appendix}, and \eqref{eq:EF_out_2_in_Appendix}, the expression inside the square brackets represents the effective light-matter interaction length within the WG for that specific Raman signal collection configuration.

% ============================================================================
\section{Raman Enhancement for Different Rhodamine 6G Modes}
\label{Appendix:EF_diff_Raman_modes}

\begin{figure}
	\centering
	\includegraphics[width=\columnwidth]{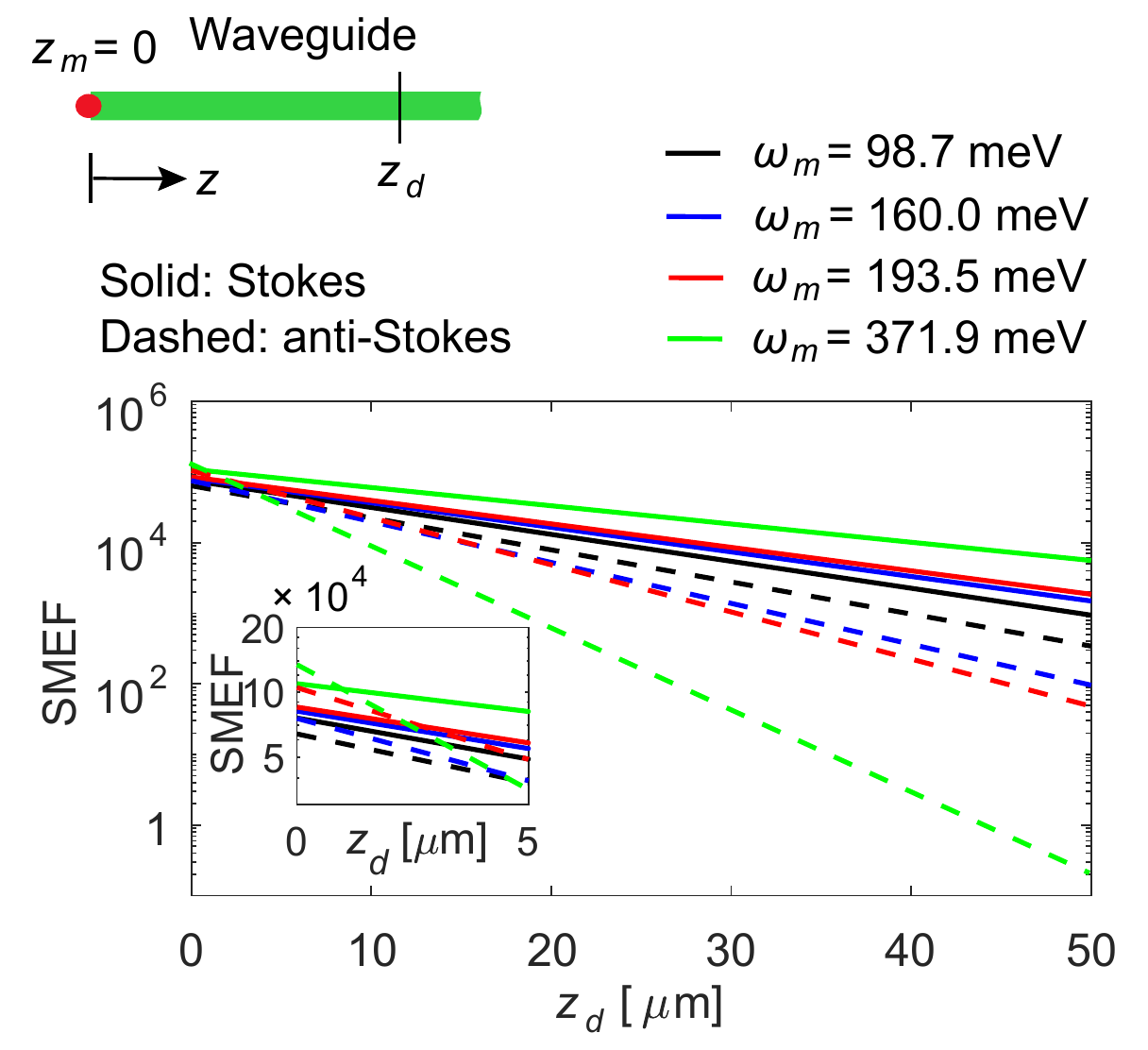}
	\caption{Single-molecule enhancement factor (SMEF) of a molecule, at the input facet ($z_m$ = 0) of an Ag plasmonic slot WG with gap width $g$ = 50 nm and also at the center of the gap region in terms of $x$-$y$ position, as a function of detection location $z_d$ along the WG, for different Raman lines. Solid: Stokes; dashed: anti-Stokes. Raman modes are indicated by black: $\omega_m$ = 98.7 meV, blue: $\omega_m$ = 160.0 meV, red: $\omega_m$ = 193.5 meV, and green: $\omega_m$ = 371.9 meV. Inset shows a magnified view of the plot for $z_d$ = 0 to 5 $\mu$m.}
	\label{fig:SMEF_Appendix}
\end{figure}

\begin{figure}
	\centering
	\includegraphics[width=\columnwidth]{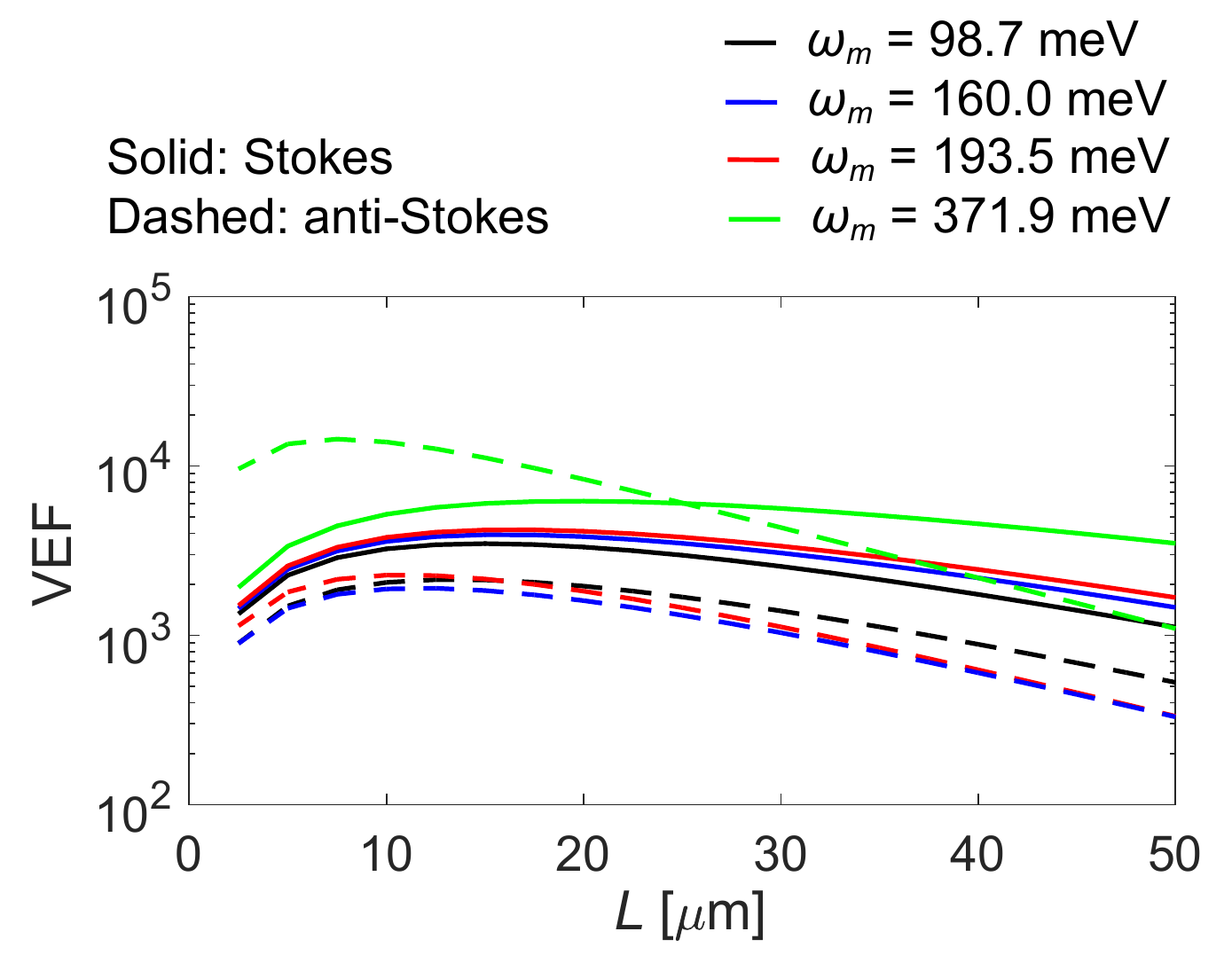}
	\caption{Volume enhancement factor (VEF) for molecules within the gap region of an Ag plasmonic slot WG with gap width $g$ = 50 nm as a function of WG length $L$ for different Raman lines (forward scattered direction). The results here are based on comparing to a reference Gaussian beam focused using an objective lens with NA = 0.75 that has a waist diameter $D_0 \approx$ 1.28 $\mu\textnormal{m}$ and depth of focus $b \approx$ 17.4 $\mu\textnormal{m}$. Note that changing the Gaussian beam parameters will simply scale the VEF. Solid: Stokes; dashed: anti-Stokes. Raman modes are indicated by black: $\omega_m$ = 98.7 meV, blue: $\omega_m$ = 160.0 meV, red: $\omega_m$ = 193.5 meV, and green: $\omega_m$ = 371.9 meV. Note that the much higher peak VEF for the anti-Stokes line of Raman mode with $\omega_m$ = 371.9 meV (dashed green) is due to the high maximum SMEF (Fig.\,\ref{fig:SMEF}(a) inset), which in turn is due to the low $\upsilon_\textnormal{E}$ [see Fig.\,\ref{fig:WGs_dispersion}(c)].}
	\label{fig:Accum_EF_Appendix}
\end{figure}

In this Appendix, we present the results of Raman enhancement by the Ag plasmonic slot WG with gap width of 50 nm in terms of both SMEF and VEF for different Raman modes of the Rhodamine 6G (R6G) molecule. Here, we use a few Raman modes of R6G as well as the C-H vibrational Raman mode as examples, each with a distinct Raman shift $\nu_m$ and the associated Raman activity $\textnormal{RA} = R_{nn}^2$. The R6G Raman modes that we utilize have shifts of $\nu_{m,1}$ = 819 $\textnormal{cm}^{-1}$ ($\omega_m$ = 98.7 meV, $\textnormal{RA}$ = 6.2 $\mathring{\textnormal{A}}^4 \textnormal{amu}^{-1}$ = 3.73 $\times 10^{-13}$ $\textnormal{m}^4\textnormal{kg}^{-1}$), $\nu_{m,2}$ = 1290.5 $\textnormal{cm}^{-1}$ ($\omega_m$ = 160.0 meV, $\textnormal{RA}$ = 5.9 $\mathring{\textnormal{A}}^4 \textnormal{amu}^{-1}$ = 3.55 $\times 10^{-13}$ $\textnormal{m}^4\textnormal{kg}^{-1}$), and $\nu_{m,3}$ = 1559 $\textnormal{cm}^{-1}$ ($\omega_m$ = 193.5 meV, $\textnormal{RA}$ = 8.2 $\mathring{\textnormal{A}}^4 \textnormal{amu}^{-1}$ = 4.94 $\times 10^{-13}$ $\textnormal{m}^4\textnormal{kg}^{-1}$) \cite{Watanabe2005}. The C-H vibrational Raman mode has shift $\nu_{m,4}$ = 3000 $\textnormal{cm}^{-1}$ ($\omega_m$ = 371.9 meV, $\textnormal{RA}$ = 7 $\mathring{\textnormal{A}}^4 \textnormal{amu}^{-1}$ = 4.22 $\times 10^{-13}$ $\textnormal{m}^4\textnormal{kg}^{-1}$). The decay rate of the Raman modes $\gamma_m$ is taken to be 1.6 meV.

We consider the scenario in which a single R6G molecule is located at the input facet of the WG at $z_m$ = 0 (also at the center of the slot region in the $x$-$y$ plane). The SMEF for Raman scattered light coupled to the WG mode decays exponentially as a function of the detection location along the WG, or equivalently, the propagation distance $z_d$ [Fig.\,\ref{fig:SMEF_Appendix}(a)]. The $z_d$ dependence is simply due to propagation (modal) loss of the WG mode carrying the Raman scattered light, which leads to exponential decay of the Raman scattered power $P^\textnormal{SM}$ as a function of $z_d$; the modal EM field distribution at the 2-D cross-section is invariant along the WG length, so only the total field amplitude or mode power decays. As seen in Fig.\,\ref{fig:SMEF_Appendix}(a), the maximal SMEF is achieved when $z_d$ = 0, which varies amongst the different Raman modes. The trends in the maximal SMEF and decay rates of SMEF as a function of $z_d$ for the Stokes and anti-Stokes lines as shown in Fig.\,\ref{fig:SMEF_Appendix}(a) are most strongly correlated with the WG dispersion rather than due to the intrinsic properties of the R6G Raman modes such as the Raman activities $\textnormal{RA} = R_{nn}^2$ and Raman shifts $\nu_m$. More specifically, it can be observed that the maximal SMEF for the Stokes lines increases as the Raman wavelength increases, which is mainly a result of the decrease in energy velocity $\upsilon_\textnormal{E}$ [Fig.\,\ref{fig:WGs_dispersion}(c)]. Similarly for the anti-Stokes lines, as the Raman wavelength is decreased, $\upsilon_\textnormal{E}$ decreases, which results in the increase in maximal SMEF as well. As seen in Fig.\,\ref{fig:SMEF_Appendix}(a), the SMEF for the Stokes lines all decay at similar rates, which is attributed to the fact that the modal losses $\alpha$ varies only over a small range [Fig.\,\ref{fig:WGs_dispersion}(b)]; whereas for the anti-Stokes lines, the larger increase in $\alpha$ as wavelength decreases causes the SMEF to decay at a faster rate as a function of $z_d$. 

In terms of volume Raman enhancement for the case of measurement on a bulk sample with an ensemble of molecules, the VEF as a function of WG length $L$ for the different R6G Raman modes are shown in Fig.\,\ref{fig:Accum_EF_Appendix}. In general, the maximal VEF at the optimal WG length $L_\textnormal{opt}$ should increase as the maximal SMEF is increased. However, note that in our calculations, the AEF taken at all spatial points in the gap region is used (Appendix \ref{Appendix:accum_enhancement}) instead of just the SMEF taken at a single spatial point (i.e., the center of the gap region). As such, the trend of the maximum VEF for the different Raman lines as shown in Fig.\,\ref{fig:Accum_EF_Appendix} would not follow exactly that of the trend in the maximal SMEF shown in Fig.\,\ref{fig:SMEF_Appendix}. Looking at the Stokes lines, as the wavelength is increased, the maximal VEF also increases, which actually follows the same trend as for the maximal SMEF (Fig.\,\ref{fig:SMEF_Appendix}). It can be seen that the optimal WG length $L_\textnormal{opt}$ that achieves the maximal VEF also increases as the wavelength is increased, which is due to the reduction in modal loss [Fig.\,\ref{fig:WGs_dispersion}(b)]. The trend in the maximal VEF for the anti-Stokes lines is less obvious, as it does not quite follow the trend in maximal SMEF shown in Fig.\,\ref{fig:SMEF_Appendix}, and this can be attributed to the point just mentioned, specifically that the VEF is calculated by utilizing the AEF within the gap region of the slot WG. For the anti-Stokes lines, $L_\textnormal{opt}$ for maximal VEF are much lower than for the Stokes lines, which follows from the higher WG modal losses experienced by the anti-Stokes lines [Fig.\,\ref{fig:WGs_dispersion}(b)].

% ======================================================================================
\section{Additional Figures}
\label{Appendix:additional_figures}

\begin{figure}[h]
	\centering
	\includegraphics[width=\columnwidth]{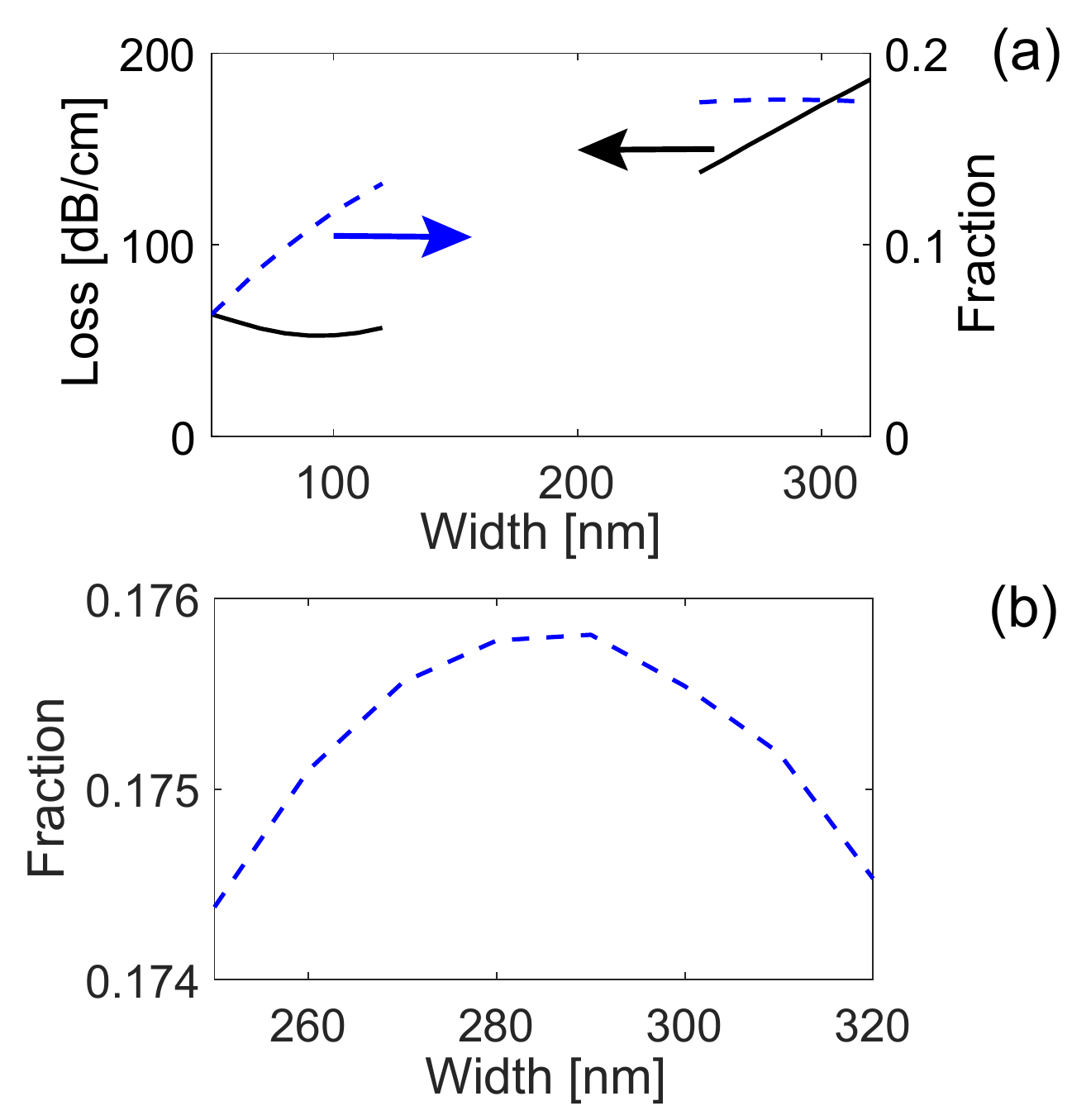}
	\caption{(a) Propagation loss and fraction of $\textbf{E}$-field intensity within gap region of the Ag-$\textnormal{Si}_3\textnormal{N}_4$ hybrid plasmonic slot WG (HPSW) as a function of the width of the left $\textnormal{Si}_3\textnormal{N}_4$ cladding at the pump wavelength of $\lambda_0$ = 785 nm. Other dimensions of the HPSW are: height = 150 nm, Ag width = 25 nm, gap width $g$ = 50 nm, and right $\textnormal{Si}_3\textnormal{N}_4$ cladding width = 250 nm. (b) Magnified plot of the fraction of $\textbf{E}$-field intensity within gap region as a function of the width of the left $\textnormal{Si}_3\textnormal{N}_4$ cladding from 250 to 320 nm.}
	\label{fig:Hybrid_wg_sweep}
\end{figure}

\begin{figure*}[h]
	\centering
	\includegraphics[width=0.95\textwidth]{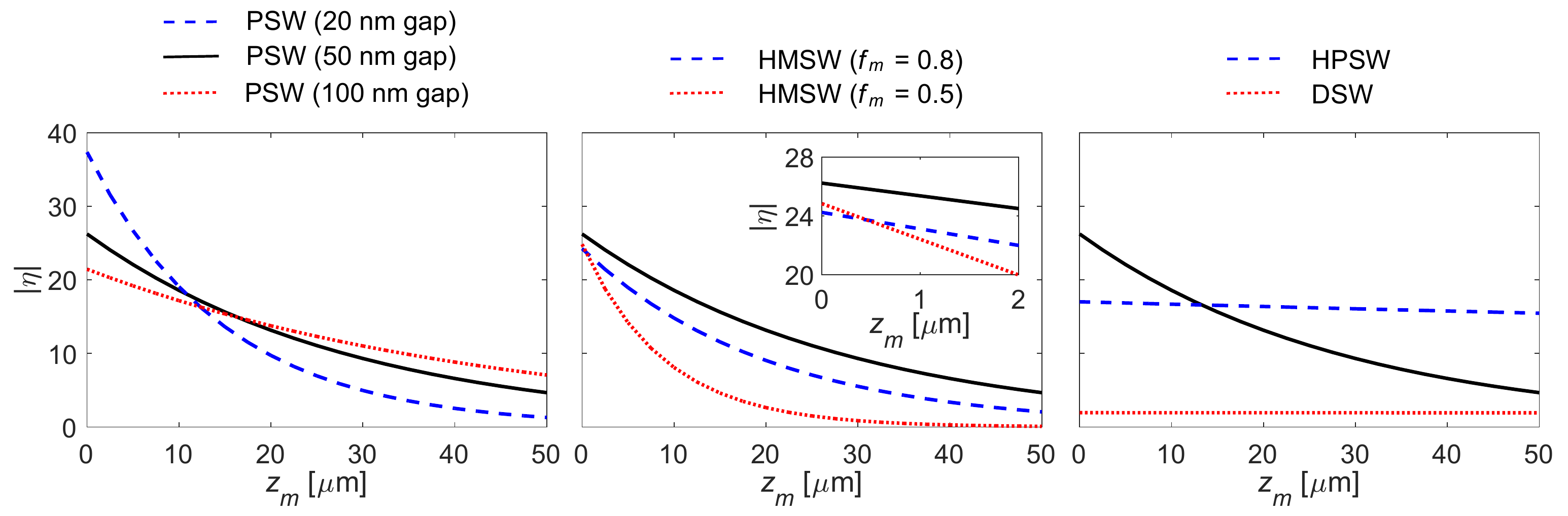}
	\caption{$\textbf{E}$-field enhancement factor $|\eta|$ of a molecule at the center of the gap region in terms of $x$-$y$ position as a function of molecule location along the WG $z_m$ at the excitation wavelength $\lambda_0$ = 785 nm for the Ag plasmonic slot WG (PSW) with gap width $g$ = 50 nm, and other WGs obtained by (a) varying the gap width $g$: Ag PSW with gap width $g$ = 100 nm and $g$ = 20 nm, (b) varying the metal filling fraction $f_m$: Ag-$\textnormal{Si}_3\textnormal{N}_4$ HM slot WG (HMSW) with $f_m$ = 0.8 and $f_m$ = 0.5, and (c) changing the WG type: Ag-$\textnormal{Si}_3\textnormal{N}_4$ hybrid plasmonic slot WG (HPSW) and $\textnormal{Si}_3\textnormal{N}_4$ dielectric slot WG (DSW). Inset of (b) shows a magnified view of the plot for $z_m$ = 0 to 2 $\mu m$.}
	\label{fig:Eta_Appendix}
\end{figure*}

\begin{figure*}[h]
	\centering
	\includegraphics[width=0.95\textwidth]{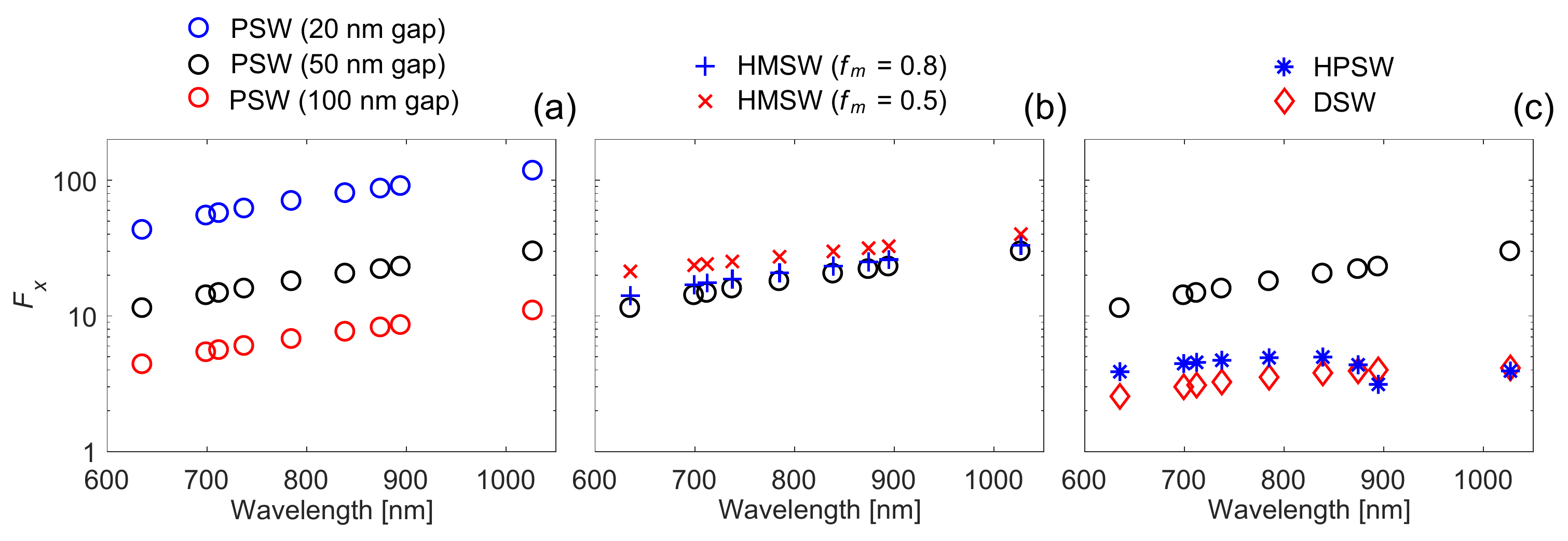}
	\caption{Purcell factor for dipole oriented in the $x$-direction $F_x$ (Eq.\,\ref{eq:Purcell_factor}) as a function of wavelength for the Ag plasmonic slot WG (PSW) with gap width $g$ = 50 nm, and other WGs obtained by (a) varying the gap width $g$: Ag PSW with gap width $g$ = 100 nm and $g$ = 20 nm, (b) varying the metal filling fraction $f_m$: Ag-$\textnormal{Si}_3\textnormal{N}_4$ HM slot WG (HMSW) with $f_m$ = 0.8 and $f_m$ = 0.5, and (c) changing the WG type: Ag-$\textnormal{Si}_3\textnormal{N}_4$ hybrid plasmonic slot WG (HPSW) and $\textnormal{Si}_3\textnormal{N}_4$ dielectric slot WG (DSW). Each marker indicates either the pump wavelength or a specific Stokes or anti-Stokes wavelength corresponding to one of the Raman modes as described in Section\,\ref{sec:WG_mode_properties}.}
	\label{fig:Fp_vs_wavelength}
\end{figure*}

\begin{figure*}[h]
	\centering
	\includegraphics[width=0.95\textwidth]{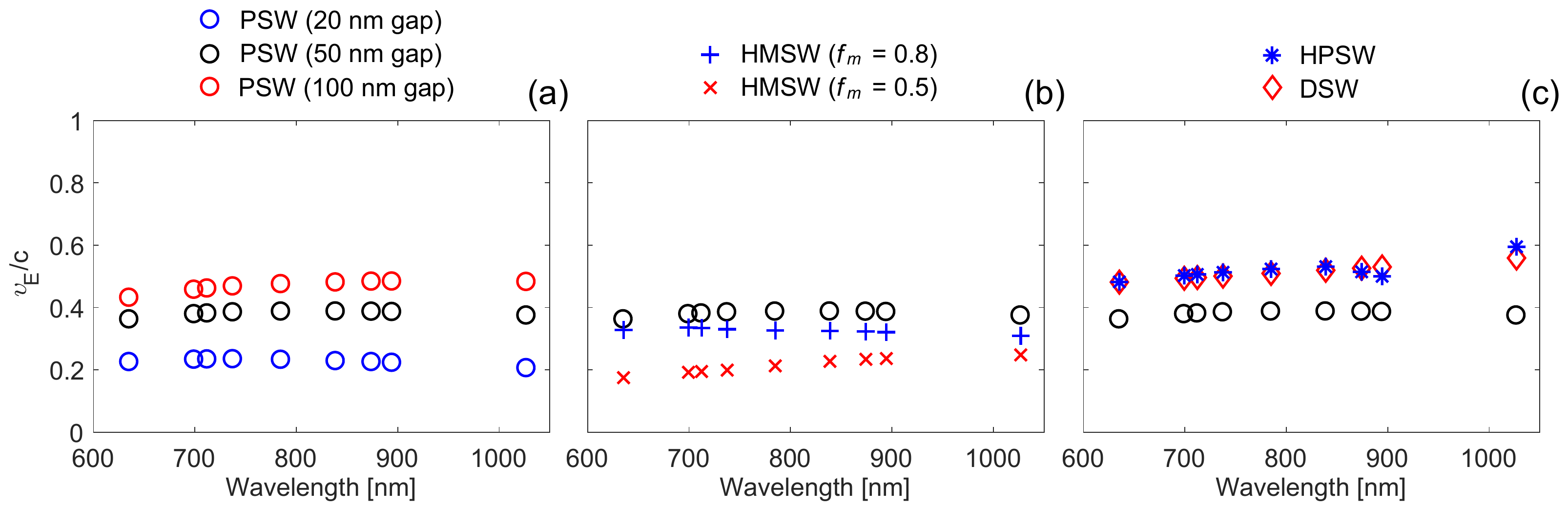}
	\caption{Normalized energy velocity $\upsilon_\textnormal{E}$/c as a function of wavelength for the Ag plasmonic slot WG (PSW) with gap width $g$ = 50 nm, and other WGs obtained by (a) varying the gap width $g$: Ag PSW with gap width $g$ = 100 nm and $g$ = 20 nm, (b) varying the metal filling fraction $f_m$: Ag-$\textnormal{Si}_3\textnormal{N}_4$ HM slot WG (HMSW) with $f_m$ = 0.8 and $f_m$ = 0.5, and (c) changing the WG type: Ag-$\textnormal{Si}_3\textnormal{N}_4$ hybrid plasmonic slot WG (HPSW) and $\textnormal{Si}_3\textnormal{N}_4$ dielectric slot WG (DSW). Each marker indicates either the pump wavelength or a specific Stokes or anti-Stokes wavelength corresponding to one of the Raman modes as described in Section\,\ref{sec:WG_mode_properties}.}
	\label{fig:Energy_velocity_vs_wavelength}
\end{figure*}

\begin{figure*}[h]
	\centering
	\includegraphics[width=\textwidth]{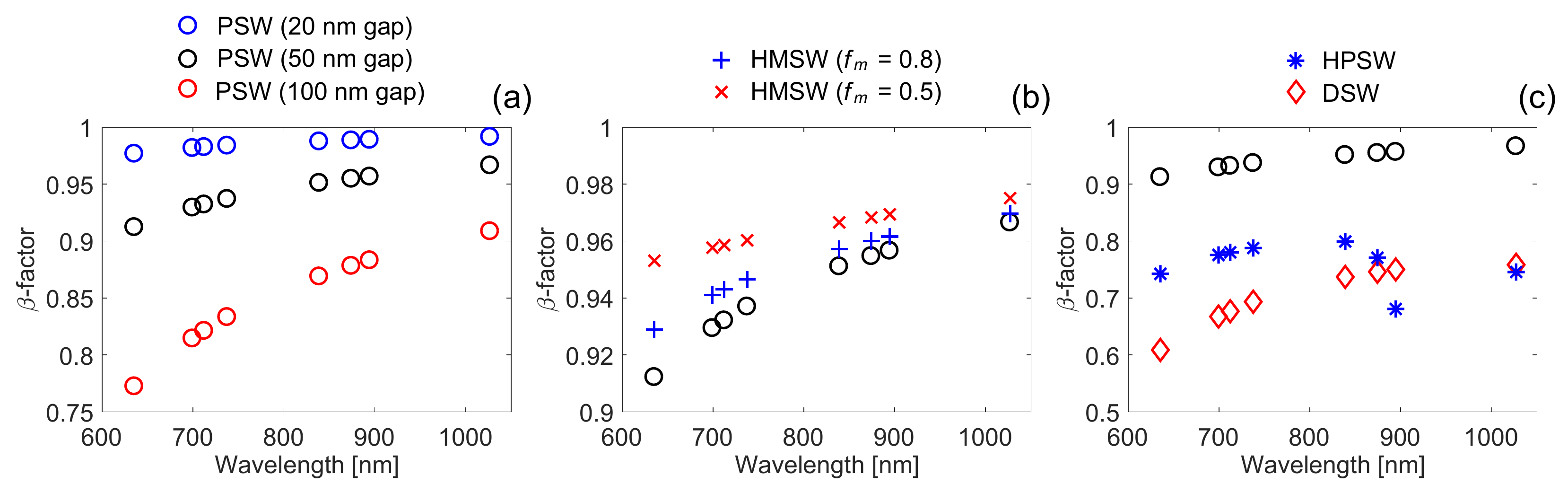}
	\caption{$\beta$-factor as a function of wavelength for the Ag plasmonic slot WG (PSW) with gap width $g$ = 50 nm, and other WGs obtained by (a) varying the gap width $g$: Ag PSW with gap width $g$ = 100 nm and $g$ = 20 nm, (b) varying the metal filling fraction $f_m$: Ag-$\textnormal{Si}_3\textnormal{N}_4$ HM slot WG (HMSW) with $f_m$ = 0.8 and $f_m$ = 0.5, and (c) changing the WG type: Ag-$\textnormal{Si}_3\textnormal{N}_4$ hybrid plasmonic slot WG (HPSW) and $\textnormal{Si}_3\textnormal{N}_4$ dielectric slot WG (DSW). Each marker indicates a specific Stokes or anti-Stokes wavelength corresponding to one of the Raman modes as described in Section\,\ref{sec:WG_mode_properties}.}
	\label{fig:beta_vs_wavelength}
\end{figure*}

\begin{figure*}[h]
	\centering
	\includegraphics[width=0.85\textwidth]{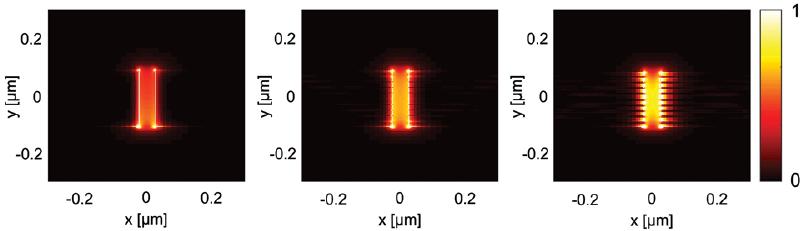}
	\caption{Electromagnetic energy density distribution $\cal{w}(\boldsymbol{\rho})$ [J/$\textnormal{m}^3$] (see Appendix\,\ref{Appendix:energy_velocity} for derivation) in $x$-$y$ cross-section of (a) Ag plasmonic slot WG, (b) Ag-$\textnormal{Si}_3\textnormal{N}_4$ HM plasmonic slot WG with $f_m$ = 0.8, and (c) Ag-$\textnormal{Si}_3\textnormal{N}_4$ HM plasmonic slot WG with $f_m$ = 0.5; each WG has gap width $g$ = 50 nm. The color bar is in linear scale (normalized units), and the plots are normalized to the same maximum \cal{w}($\boldsymbol{\rho}$).}
	\label{fig:EM_energy_dens}
\end{figure*}

\begin{figure}[h]
	\centering
	\includegraphics[width=0.8\columnwidth]{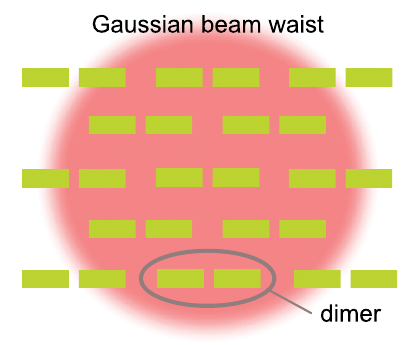}
	\caption{Schematic of Gaussian beam waist on a SERS substrate surface with an array of plasmonic dimers; only $\sim$10 dimers can be excited by the Gaussian beam with waist diameter $\sim$638 nm.}
	\label{fig:beam_excitation_SERS_dimers}
\end{figure}

\clearpage

\bibliographystyle{apsrev4-1}
\bibliography{WG_SERS_REF_v13}

%\printbibliography

\end{document}